\documentclass[lettersize,journal]{IEEEtran}
\usepackage{soul}
\usepackage{url}
\usepackage{multirow}
\usepackage[utf8]{inputenc}
\usepackage[font=small]{caption}
\usepackage{graphicx}
\usepackage{amsmath}
\usepackage{amsfonts}
\usepackage{amsthm}
\usepackage{booktabs}
\usepackage{algorithm}
\usepackage{algorithmic}
\usepackage[switch]{lineno}
\usepackage{csquotes}
\usepackage{bm}
\usepackage[font=small]{caption}
\usepackage{array}
\usepackage{cite}

\usepackage{array}  
\usepackage{threeparttable,multirow,booktabs}
\usepackage[table,xcdraw]{xcolor}

\newcommand{\Black}[1]{{\color{black}{#1}}}

\definecolor{bestblue}{RGB}{205,227,255}
\definecolor{secondblue}{RGB}{243,249,255}
\definecolor{abred1}{RGB}{244,156,156}   
\definecolor{abred2}{RGB}{247,186,186}   
\definecolor{abred3}{RGB}{250,214,214}   
\definecolor{abred4}{RGB}{252,234,234}   
\definecolor{abred5}{RGB}{254,246,246}   

\usepackage{todonotes}

\usepackage{bm}
\usepackage{comment}
\begin{document}

\title{CDNet: Combined Dictionary Unfolding Network for Parameter-Efficient Multi-Source Fusion}

\title{Combined Dictionary Unfolding Network with Gradient-Adaptive Fidelity for \Black{Transferable} Multi-Source Fusion}

\author{
Ge~Luo$^{\dagger}$,
Jun-Jie~Huang$^{\dagger}$,
Qi~Yu,
Tianrui~Liu,
Ke~Liang, 
Yuming~Xiang,~\IEEEmembership{Senior Member,~IEEE},
Wentao~Zhao,
Xinwang~Liu,~\IEEEmembership{Senior Member,~IEEE},
and~Meng~Wang,~\IEEEmembership{Fellow,~IEEE}
\thanks{G. Luo, J.-J. Huang, Q. Yu, T. Liu, W. Zhao and X. Liu are with the College of Computer Science and Technology, National University of Defense Technology, Changsha, China (email: luoge25@nudt.edu.cn; jjhuang@nudt.edu.cn; yuqi10@nudt.edu.cn; trliu@nudt.edu.cn; liangke200694@126.com; wtzhao@nudt.edu.cn; xinwangliu@nudt.edu.cn). Y. Xiang is with College of Surveying and Geoinformatics, Tongji University, Shanghai, China (email: xiangym@tongji.edu.cn). M. Wang is with the School of Computer and Information, Hefei University of Technology, Hefei 230009, China (email: eric.mengwang@gmail.com). $^{\dagger}$ denotes Equal contributions.}

}

\markboth{Submitted to IEEE Transactions on Pattern Analysis and Machine Intelligence}%
{Shell \MakeLowercase{\textit{et al.}}: A Sample Article Using IEEEtran.cls for IEEE Journals}

\maketitle

\begin{abstract}

Deep Unfolding Network (DUN)-based methods have emerged as effective solutions for multi-source image fusion by combining model-driven iterative optimization with data-driven deep learning. However, most existing deep unfolding image fusion methods are derived from alternating minimization, which updates the features of different modalities separately. This design introduces considerable computational and memory overhead, limiting deployment on resource-constrained edge devices. \Black{To address this issue, we propose CDNet, a lightweight Combined Dictionary Unfolding Network for multi-source image fusion. 
Rather than introducing a new sparse coding prior or empirically compressing an existing fusion network, CDNet translates the unique-common decomposition prior of coupled dictionary learning into a structurally constrained joint unfolding architecture. 
The resulting CDBlock follows a block-sparse interaction topology and performs a model-derived joint update of common and modality-specific representations, thereby streamlining feature learning and improving efficiency.} 
In addition, we design a compact High- and Low-frequency Image Fidelity (HLIF) loss for unsupervised training without ground-truth images.
We evaluate CDNet on four tasks, including multi-exposure image fusion, infrared and visible image fusion, medical image fusion, and infrared and visible image fusion for semantic segmentation. \Black{Although trained only on the SICE dataset, CDNet shows promising transferability across different fusion scenarios.} Experimental results show that CDNet achieves competitive or superior fusion performance with high efficiency. For infrared and visible image fusion, CDNet outperforms competing methods on four of six metrics on the TNO dataset and five of six metrics on the RoadScene dataset. In particular, it surpasses the second-best method by 1.23 dB and 1.59 dB in PSNR on TNO and RoadScene, respectively. 

\end{abstract}


\section{Introduction}

\IEEEPARstart{M}{ulti-Source} Image Fusion (MSIF)~\cite{stathaki2011image,wu2025fully} is a key task in computer vision that integrates complementary information from heterogeneous image sources, \textit{e.g.}, infrared-visible, multi-exposure, or multi-modal medical images, to generate a more informative composite image for visual perception and downstream high-level tasks. In practical applications, MSIF methods are often expected to run on edge devices with limited computational power, memory, and energy. Therefore, lightweight and high-efficiency fusion models are important for robust deployment under strict resource budgets. Beyond efficiency, practical MSIF systems also require cross-modal transferability, namely the ability to train on one modality pair (\textit{e.g.}, multi-exposure RGB) and deploy on heterogeneous fusion tasks (\textit{e.g.}, medical or thermal fusion) without task-specific retraining or fine-tuning.

\begin{figure}[]
  \centering
  \includegraphics[width=0.48\textwidth]{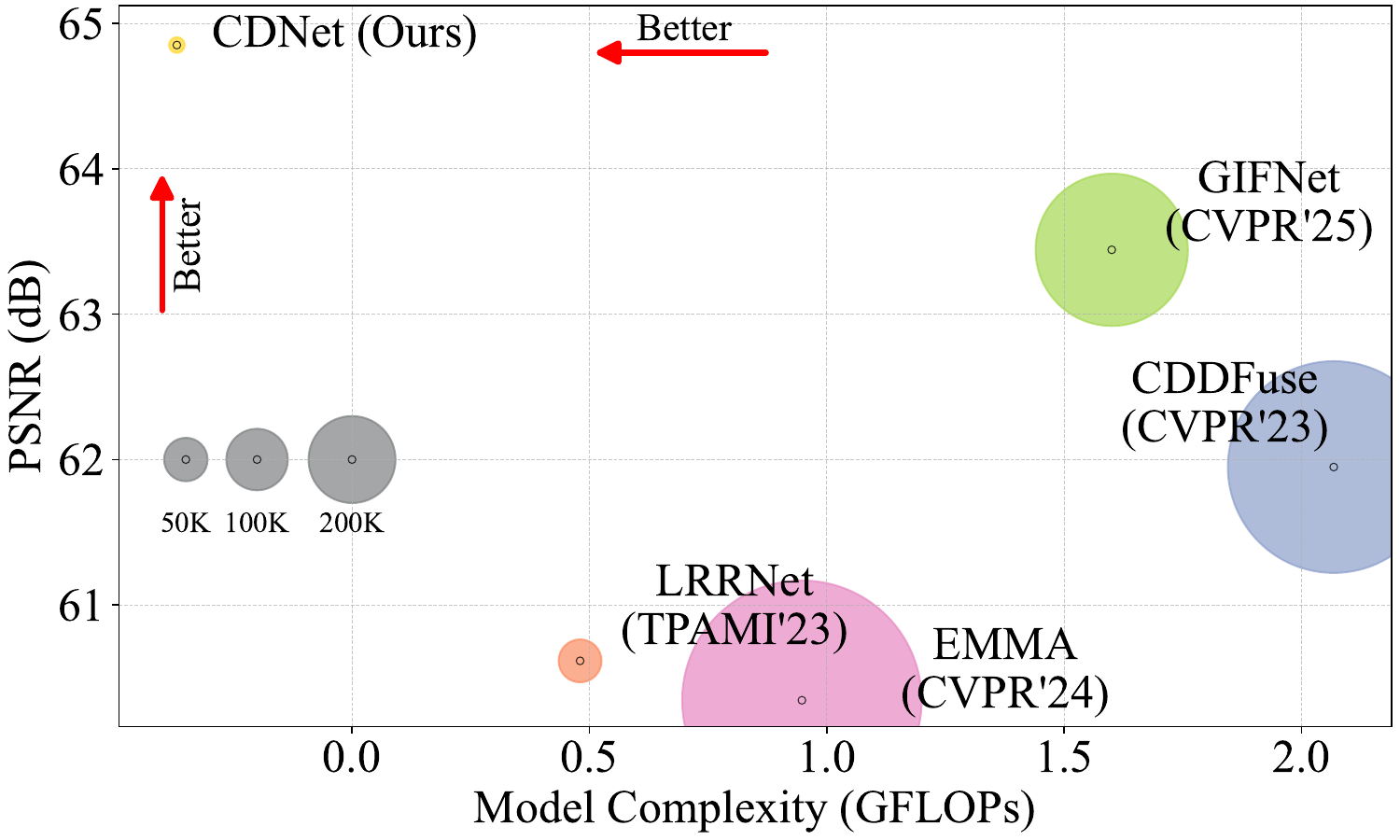}
  \caption{Comparison of leading image fusion methods on TNO and RoadScene datasets. The results are evaluated based on average PSNR (dB), number of parameters (M) and complexity (GFLOPs). Note: GFLOPs values are shown on a log10 scale. The model size is depicted as the area of the ball.}
  \label{fig:comparison}
\end{figure}

Early MSIF methods relied on handcrafted transforms, such as Intensity-Hue-Saturation (IHS)~\cite{tu2001efficient, choi2006new} and Laplacian pyramid~\cite{burt1983multiresolution, du2016union, wang2020multi, yao2023laplacian}, but their adaptability to complex imaging conditions is limited. With the rise of deep learning, Deep Neural Network (DNN)-based methods have achieved remarkable fusion performance by learning data-driven feature representations~\cite{ram2017deepfuse, liu2017multi, li2018densefuse, liu2018deep, wang2019enhanced, liu2020multi}. However, many of them behave as black-box mappings and often require large model capacity. To bridge model-driven optimization and data-driven learning, Deep Unfolding Network (DUN)-based methods~\cite{deng2020deep,deng2023deepm,li2023lrrnet,fang2024deep,bai2024deep,he2023degradation} unfold iterative optimization procedures into learnable network architectures, providing a promising balance between interpretability and representation ability.

Despite their promise, prevailing DUN-based MSIF methods still face two important limitations. First, their feature update strategies are often inefficient. Most DUN-based methods~\cite{deng2020deep,gao2022multi,li2023lrrnet} are derived from coupled dictionary learning, where source images are decomposed into multiple components, such as common and unique features, each associated with a dedicated dictionary. Solving the resulting optimization problem usually requires alternating updates over different representations. When unfolded into a network, this strategy leads to separate branches, sequential computation, and additional memory access, making the model less compact and less friendly to edge deployment. Second, many unsupervised fusion losses combine multiple heterogeneous objectives, including metric-based fidelity (\textit{e.g.}, SSIM and MSE)~\cite{tang2022ydtr,cheng2025one}, gradient preservation~\cite{zheng2023efficient}, perceptual alignment with pre-trained networks~\cite{li2023lrrnet,tang2025mask}, and even modality- or task-specific constraints~\cite{tang2022image, cheng2025one, tang2022ydtr}. Although effective, these multi-term formulations require careful balancing and may complicate optimization.

\Black{To address these challenges, we propose CDNet, a lightweight Combined Dictionary Unfolding Network for MSIF. 
Rather than compressing an existing fusion network or introducing a new sparse coding prior, CDNet translates the coupled dictionary prior into a structurally constrained joint unfolding architecture. 
The common and modality-specific representations are organized under a block-sparse interaction topology, enabling CDBlock to perform a model-derived joint update instead of fragmented alternating updates. 
We further introduce a compact High- and Low-frequency Image Fidelity (HLIF) loss, which constructs gradient-adaptive structural and luminance references for unsupervised and transferable fusion.}

\Black{Fig.~\ref{fig:comparison} compares representative MSIF methods in terms of average PSNR, parameter count, and computational complexity on the TNO and RoadScene datasets. CDNet achieves a favorable accuracy-efficiency trade-off, using only about 1/94 of the parameters and 1/93 of the computational complexity required by GIFNet, the second-best method in PSNR}~\cite{cheng2025one}.

The main contributions of this paper are summarized as follows:

\Black{
\begin{itemize}
\item We translate the common/source-specific decomposition prior of coupled dictionary learning into a block-sparse joint representation for structured unfolding, enabling all representation components to be updated within a unified step.

\item We derive CDBlock as a structurally constrained convolutional realization of the joint proximal-gradient update, replacing fragmented alternating branches with a model-derived lightweight fusion block.

\item We introduce a compact HLIF loss with gradient-adaptive high- and low-frequency references. Experiments across MEF, IVF, MIF, and downstream segmentation demonstrate a favorable balance between fusion quality, efficiency, and cross-task generalization.
\end{itemize}
}

The remainder of this paper is organized as follows. Section~\ref{sec:related} reviews related work. Section~\ref{sec:CDict} introduces the coupled dictionary model. Section~\ref{sec:proposed} presents the proposed CDNet. Section~\ref{sec:complexity} analyzes computational efficiency. Section~\ref{sec:experiment} reports experimental results. Finally, Section~\ref{sec:conclusion} concludes this paper.

\section{Related Works}
\label{sec:related}

This section reviews three groups of studies closely related to our work, including deep neural network-based fusion methods, deep unfolding-based fusion methods, and loss functions for unsupervised image fusion.

\subsection{DNN-based Fusion Methods}
\label{sec:dnns}

Deep Neural Network (DNN)-based fusion methods have achieved remarkable progress by learning data-driven feature representations from paired images~\cite{ram2017deepfuse, li2018densefuse, amin2019ensemble, zhang2020ifcnn, xu2020u2fusion, zhang2020rethinking, zhang2021image}. Early autoencoder-based methods~\cite{ram2017deepfuse, li2018densefuse} usually adopt convolutional encode-fuse-decode pipelines, where source features are first extracted, then fused, and finally reconstructed into the fused image. Although effective, these methods are often designed as data-driven black-box models and lack explicit mechanisms to explain why specific source features are preserved or suppressed. Therefore, they usually rely on increased architectural capacity to improve representation ability. For example, DenseFuse~\cite{li2018densefuse} introduces dense blocks~\cite{huang2017densely} into the encoder, while ECNN~\cite{amin2019ensemble} employs multiple CNNs to improve decision map generation.

Recent studies further employ Transformer architectures~\cite{ma2022swinfusion,tang2022ydtr, cheng2025one, MSRQ} and diffusion models~\cite{huang2024fusiondiff,tang2024drmf} to enhance cross-modal interaction and restoration ability. For instance, GIFNet~\cite{cheng2025one} incorporates Swin Transformer blocks for feature interaction, while DRMF~\cite{tang2024drmf} adopts iterative diffusion sampling for degradation-robust fusion. These advanced architectures improve the representation capacity of fusion models, but they also introduce additional computational cost. Self-attention has quadratic complexity with respect to image size, and diffusion-based models require repeated sampling steps during inference. Therefore, how to achieve a favorable balance among fusion performance, interpretability, and computational efficiency remains an important issue for practical MSIF deployment.

\subsection{Deep Unfolding-based Fusion Methods}
\label{sec:duns}

Deep Unfolding Network (DUN)-based fusion methods bridge model-driven iterative optimization and data-driven deep learning by mapping each iteration of an optimization algorithm to a network layer~\cite{daubechies2004iterative,gregor2010learning,zhang2018ista}. Through this unfolding mechanism, manually designed parameters in traditional models, such as regularization coefficients, dictionary atoms, and thresholding operators, become learnable network parameters optimized end-to-end. Therefore, DUNs combine the interpretability of iterative algorithms with the representation capacity of deep neural networks~\cite{xie2020mhf,pu2022mixed,marivani2022designing,he2023degradation,huang2025lightweight,DFDUN}.

Several studies introduce sparse representation and dictionary learning into deep unfolding frameworks for image fusion. MCDL~\cite{gao2022multi} and DeepM\textsuperscript{2}CDL~\cite{deng2023deepm} unfold coupled dictionary learning models with shared and modality-specific components, while DeRUN~\cite{he2023degradation} embeds structural priors into a robust unfolding framework. CSCFuse~\cite{zhao2023deep}, CCSR-Net/MCCSR-Net~\cite{zheng2025unfolding}, and FNet~\cite{panda2024l0} further explore interpretable sparse coding-based fusion models under different formulations. These methods demonstrate that deep unfolding provides an effective paradigm for improving model transparency while maintaining adaptability across different fusion tasks.

The efficiency of DUN-based fusion models is closely related to the number of dictionaries, representations, and update branches involved~\cite{deng2020deep,gao2022multi,li2023lrrnet}. For example, CU-Net~\cite{deng2020deep} alternately updates three feature representations associated with common and unique components, resulting in a relatively complex update procedure. LRRNet~\cite{li2023lrrnet} improves efficiency by integrating low-rank and sparse priors into two combined representations, but it still relies on alternating updates. In parallel, some recent works attempt to avoid explicit component decomposition~\cite{yin2015sparse,zhao2023deep,zheng2025unfolding}. CSCFuse~\cite{zhao2023deep} learns unified feature representations through interpretable convolutional dictionary units, while CCSR-Net~\cite{zheng2025unfolding} introduces a coupled updating scheme for bidirectional interaction between source representations. \Black{In contrast, CDNet preserves the common/source-specific prior of coupled dictionary learning, but changes its architectural realization from branch-wise alternating updates to a block-sparse joint unfolding block. 
This distinguishes CDNet from methods that either remove explicit component decomposition or unfold component-wise alternating updates.}

\subsection{Loss Function Design}
\label{sec:lossdesign}

Due to the absence of ground-truth fused images, unsupervised loss design plays a crucial role in image fusion. Metric-based losses are among the most widely used objectives. They usually compute quantitative similarity between source and fused images, such as SSIM and MSE, and combine these metrics to guide model training~\cite{raza2020pfaf, jian2020sedrfuse, liu2021learning, wang2022res2fusion, liu2024task}. However, these losses mainly emphasize pixel-level similarity and may not sufficiently capture perceptual or structural consistency.

To improve perceptual quality, perception-based losses have been introduced. They usually rely on pre-trained networks to extract multi-level features from source and fused images, and then minimize their feature discrepancies~\cite{zhang2020ifcnn, zhao2024equivariant, cheng2025one, tang2025mask}. For example, LRRNet~\cite{li2023lrrnet} uses a pre-trained VGG-16 network to align shallow, middle, and deep features, while EMMA~\cite{zhao2024equivariant} employs two U-Nets to reconstruct pseudo source pairs for loss computation. Mask-DiFuser~\cite{tang2025mask} also aligns the fused image with a pseudo ground-truth through VGG-based perceptual supervision. Although perception-based losses can enhance semantic consistency, they depend on external pre-trained models and introduce additional computational overhead.

Gradient-based losses focus on structural and textural fidelity by constraining gradient differences between the fused and source images~\cite{zhang2020rethinking, cheng2023mufusion, cao2023multi, zheng2023efficient}. For instance, PMGI~\cite{zhang2020rethinking} and MoE-Fusion~\cite{cao2023multi} use gradient magnitude to preserve salient edges, while FFMEF~\cite{zheng2023efficient} further introduces a binary mask to balance high- and low-frequency fusion. These losses are effective in preserving fine details and edge structures, but they may be sensitive to noise and are less effective in maintaining global luminance consistency. \Black{Motivated by these observations, we design a compact unsupervised loss that uses gradient-adaptive weights to construct separate high-frequency and low-frequency references, improving structural preservation while maintaining luminance consistency.}
\section{Preliminary: Coupled Dictionary Model}
\label{sec:CDict}

Coupled dictionary learning \cite{yang2012coupled,guo2014online,deng2019deep,veshki2021coupled} serves as a foundational framework for multi-modal image fusion,
leveraging coupled dictionaries and joint sparse representations to fuse information from heterogeneous image modalities into a single and more informative image. 
By explicitly modeling the common feature and unique features, coupled dictionary learning enables high-fidelity fusion that preserves both the consistent scene structure and the unique modality characteristics.

For a pair of source images $\bm{x}, \bm{y} \in \mathbb{R}^{H \times W}$, they are first passed a feature extraction module to obtain representations $\bm{X}, \bm{Y} \in \mathbb{R}^{C \times H \times W}$. 
Then, they are represented by a unique sparse representation encoding modality-specific information and a common sparse representation encoding cross-modal structural information over corresponding dictionaries:
\begin{equation}
\label{eq:dict}
    \begin{cases}
        \bm{X}&=\bm{U_X} \bm{Z_X}+\bm{C_{X}} \bm{Z_C},\\
        \bm{Y}&=\bm{U_Y} \bm{Z_Y}+\bm{C_{Y}} \bm{Z_C},
    \end{cases}
\end{equation}
where $\bm{Z}_{\bm{X}}, \bm{Z}_{\bm{Y}}$ denote the unique feature representations of $\bm{X}$ and $\bm{Y}$, respectively, and $\bm{Z_C}$ represents their common feature representation. Moreover, $\bm{U}_{\bm{X}}, \bm{U}_{\bm{Y}}$ are the dictionaries corresponding to the unique features, and $\bm{C}_{\bm{X}}, \bm{C}_{\bm{Y}}$ are the common dictionaries associated with the common feature representations, respectively.

By imposing a sparsity prior on the representations, we can recast the common and unique feature decomposition problem as a more tractable sparse coding problem over the dictionaries:
\begin{equation}
\begin{aligned}
\arg\underset{\{\bm{Z}_I\}_{I\in\{\bm{X,Y,C}\}}}{\min} &\frac{1}{2} ( \left\|\bm{X} - \bm{U_X}\bm{Z_X} - \bm{C_{X}}\bm{Z_C} \right\|_F^2 \\ & + \left\|\bm{Y} - \bm{U_Y}\bm{Z_Y} - \bm{C_{Y}}\bm{Z_C} \right\|_F^2 ) \\ & +  \lambda  \sum_{I \in \{\bm{X}, \bm{Y}, \bm{C}\}} \left\|\bm{Z}_I \right\|_1,
\end{aligned}
\label{eq:update}
\end{equation}
{where $\lambda$ is a regularization parameter, and $\Vert \cdot \Vert_F$ and $\Vert \cdot \Vert_1$ denote the Frobenius norm and $\mathcal{L}_1$ norm, respectively.}

\textbf{Optimization Strategy:}
This coupled sparse coding problem involves jointly optimizing three independent sparse variables $\{\bm{Z}_I\}_{I\in\{\bm{X,Y,C}\}}$. 
Standard optimization employs an alternating minimization strategy. In each iteration, one variable is updated while the others are fixed, cycling through all variables. When unfolded into a deep network, each iteration becomes a neural network layer, and each variable update requires dedicated computational branches.

Specifically, for a single proximal gradient step updating $\bm{Z_X}$ (with $\bm{Z_Y, Z_C}$ fixed), the computation involves: two dictionary-feature multiplications for gradient computation $\bm{U_X}^{\text{T}}\bm{(U_XZ_X - X)}$, two additional multiplications for the common component coupling $\bm{C_X^{\text{T}}(C_XZ_C - X)}$ and one nonlinear proximal operator (soft-thresholding) for sparsity enforcement.
This pattern repeats for $\bm{Z_Y}$ and $\bm{Z_C}$ updates, resulting in sequential, non-parallelizable computation.
Notably, we assume that the soft-thresholding operator is used as the proximal operator, but if a more expressive neural network is adopted to approximate the proximal operator, the computational overhead is further amplified.

\textbf{Motivation:}
The analysis reveals that the efficiency bottleneck stems not from the coupled representation itself, but from the multiplicative nature of alternating updates. Each additional modality or decomposed component linearly increases both parameter count and computational latency. \Black{This observation motivates us to keep the coupled dictionary prior unchanged, but to change how this prior is unfolded into a network. 
Instead of implementing the common and source-specific updates as separate alternating branches, we map their block-wise coupling structure to a structurally constrained joint unfolding block. 
Thus, the novelty lies in the optimization-to-architecture translation of the coupled prior, rather than in replacing the underlying sparse coding model.}

\section{The Proposed Method}
\label{sec:proposed}
\begin{figure*}[t]
  \centering
  \includegraphics[width=0.92\textwidth]{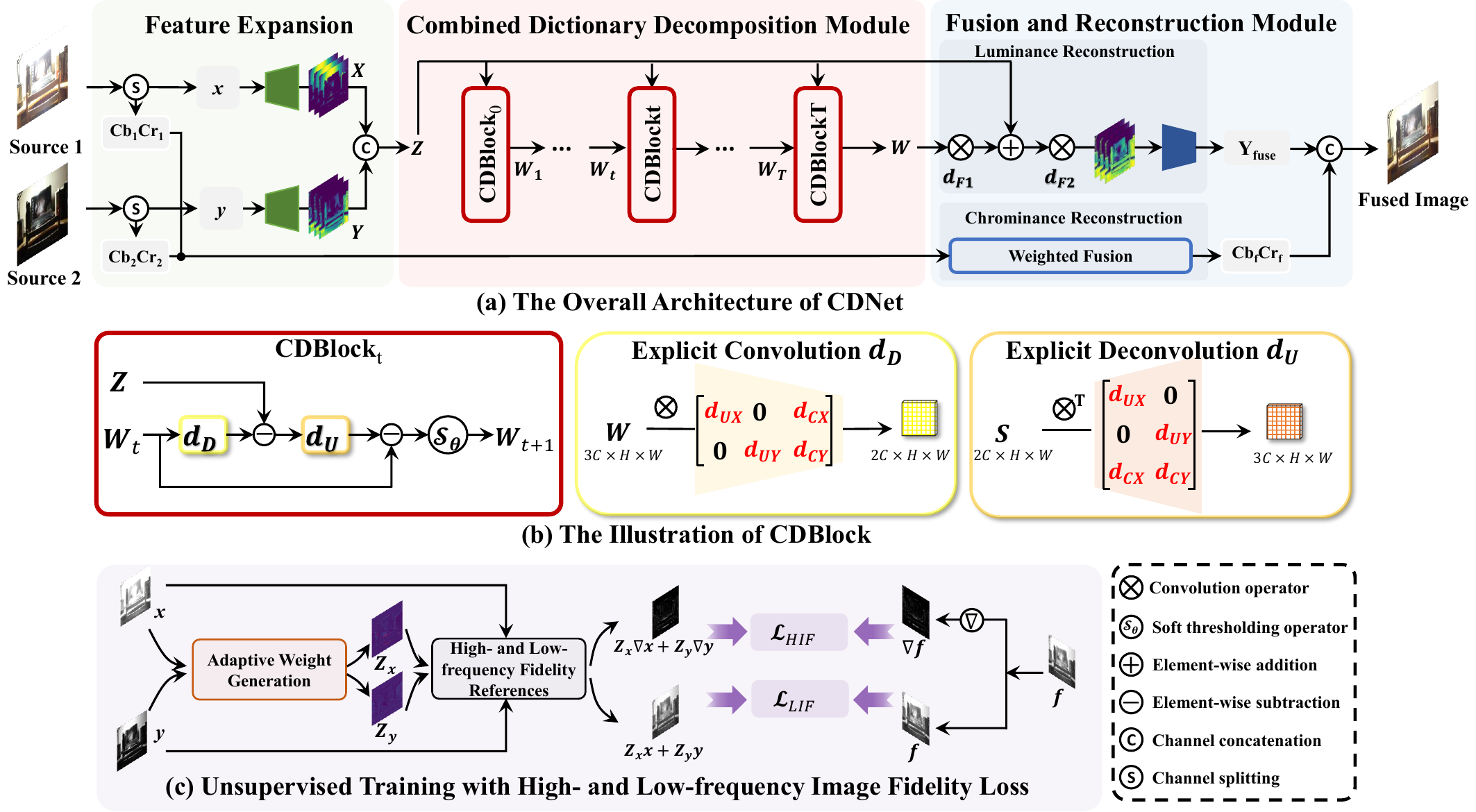}
 
    \caption{Workflow of CDNet. The Y-channel inputs are first expanded and concatenated as $\bm{Z}$, then refined by iterative CDBlocks into the unified representation $\bm{W}$. \Black{Each CDBlock is derived from the LISTA-style update of the combined sparse coding objective, where $\bm{d_D}$ and $\bm{d_U}$ follow the block-structured topology of $\bm{D}$ and $\bm{D}^{\text{T}}$ to jointly update source-specific and common representations.} The fused Y-channel is reconstructed from $\bm{W}$ with a residual connection from $\bm{Z}$, and then guides chrominance fusion. The bottom inset shows the proposed HLIF loss, where adaptive weight generation produces $\bm{Z_x}$ and $\bm{Z_y}$ to construct weighted high- and low-frequency references for supervising $\nabla \bm{f}$ and $\bm{f}$ through $\mathcal{L}_{HIF}$ and $\mathcal{L}_{LIF}$, respectively.}
    
  \label{fig:method}
\end{figure*}

This section details the proposed CDNet, a lightweight Combined Dictionary Unfolding Network for Multi-Source Image Fusion (MSIF). 
It is designed to resolve the architectural fragmentation while simultaneously enabling promising cross-task transferability.

\subsection{Overall Architecture}

Departing from the alternating implementation of coupled dictionary learning in Eq.~(\ref{eq:update}), CDNet keeps the unique-common decomposition prior but changes its unfolding architecture: all representation components are refined through a structurally constrained joint update rather than through separate alternating branches.

To achieve cross-task transferability, CDNet operates exclusively on the luminance (Y) channel in YCbCr color space. This design is {motivated by the empirical observation} that the Y channel represents intensity information, which is shared across different imaging sensors, and chrominance channels may introduce task- or sensor-specific variations that reduce cross-task transferability. \Black{We further design a compact High- and Low-frequency Image Fidelity loss, which provides modality-agnostic gradient-based supervision and improves cross-task generalization.}

As in Fig.~\ref{fig:method}, the CDNet comprises three stages:
(i) Feature Expansion: expanding input Y channels $\bm{x}, \bm{y} \in \mathbb{R}^{H \times W}$ to features $\bm{X}, \bm{Y} \in \mathbb{R}^{C\times H \times W}$. (ii) Unfolded Combined Dictionary Decomposition Module iteratively refines a unified representation $\bm{W}$ with CDBlocks. (iii) Fusion and Reconstruction Module projects $\bm{W}$ back to the fused Y channel, which is combined with chrominance via weighted fusion. The chrominance channels are then fused using weighted fusion~\cite{ram2017deepfuse, xu2020u2fusion, zheng2023efficient}. In addition to the inference pipeline, Fig.~\ref{fig:method} also summarizes the proposed HLIF loss used during training, where adaptive weights are generated from source gradients to construct high- and low-frequency references for supervising the fused Y-channel.
For clarity, the main notations used in this section are listed in Table \ref{tab:notations}.

\subsection{\Black{From Coupled Dictionary Prior to Structurally Constrained CDBlock}}
\label{sec:CDM}

\Black{CDBlock translates the coupled dictionary prior into a structurally constrained joint unfolding block, where common and modality-specific representations are updated synchronously under a block-sparse interaction topology.}

\subsubsection{\Black{Joint Representation for Structured Unfolding}}

Traditional methods require learning separate dictionaries $\{\bm{U_X},\bm{C_X},\bm{U_Y}, \bm{C_Y}\}$ and employing alternating updates for three variables $\{\bm{Z_X},\bm{Z_Y},\bm{Z_C}\}$.
For the purpose of joint unfolding, we reorganize these variables by defining a combined feature representation $\bm{W}=\left[\bm{Z_X},\bm{Z_Y},\bm{Z_C}\right]^{\text{T}}$ and a combined input $\bm{Z}=\left[\bm{X};\bm{Y}\right]^T$. 
The coupled dictionaries are consolidated into a single block-structured combined dictionary $\bm{D}$:
\begin{equation}
    \bm{D}=\left[\begin{matrix}\bm{U_X}&\bm{0}&\bm{C_{X}}\\\bm{0}&\bm{U_Y}&\bm{C_{Y}}\\\end{matrix}\right],
\end{equation}
where $\bm{0}$ denotes an all-zeros matrix. This block structure explicitly encodes the prior-guided \enquote{unique-common} separation while enabling joint optimization.

\begin{table}[t]
\centering
\renewcommand{\arraystretch}{1.2}
\caption{The list of main notations used in this paper.}
\resizebox{0.5\textwidth}{!}{
\begin{tabular}{m{1.5cm}<{\raggedright\arraybackslash} m{1.6cm}<{\raggedright\arraybackslash} m{4cm}<{\raggedright\arraybackslash}}
\toprule[1pt]
Notation & Dimension & Description \\ \midrule \midrule
$\bm{x}$, $\bm{y}$ & $\mathbb{R}^{\scriptscriptstyle H\times W}$ & Input Y channels (YCbCr). \\
$\bm{f}$ & $\mathbb{R}^{\scriptscriptstyle H\times W}$ & Fused Y channel (YCbCr). \\
$\bm{X}$, $\bm{Y}$ & $\mathbb{R}^{\scriptscriptstyle C\times H\times W}$ & Expanded features. \\
$\bm{F}$ & $\mathbb{R}^{\scriptscriptstyle C\times H\times W}$ & Multi-channel fused feature. \\
$\bm{d}_{\scriptscriptstyle \bm{UX}}$, $\bm{d}_{\scriptscriptstyle \bm{UY}}$ & $\mathbb{R}^{\scriptscriptstyle C\times C \times s\times s}$ & Unique dictionary kernels.\\
$\bm{d}_{\scriptscriptstyle \bm{CX}}$, $\bm{d}_{\scriptscriptstyle \bm{CY}}$ & $\mathbb{R}^{\scriptscriptstyle C\times C \times s\times s}$ & Common dictionary kernels. \\
$\bm{d}_{\scriptscriptstyle \bm{U}}$ & $\mathbb{R}^{\scriptscriptstyle 3C\times 2C \times s\times s}$ & Explicit deconvolution operator. \\
$\bm{d}_{\scriptscriptstyle \bm{D}}$ & $\mathbb{R}^{\scriptscriptstyle 2C\times 3C \times s\times s}$ & Explicit convolution operator. \\
$\bm{W}_{\scriptscriptstyle \bm{X}}$, $\bm{W}_{\scriptscriptstyle \bm{Y}}$ & $\mathbb{R}^{\scriptscriptstyle C\times H\times W}$ & Unique features of $\bm{X}$ and $\bm{Y}$. \\
$\bm{W}_{\scriptscriptstyle \bm{C}}$ & $\mathbb{R}^{\scriptscriptstyle C\times H\times W}$ & Common feature representation. \\
\bottomrule[1pt]
\end{tabular}
}
\label{tab:notations}
\end{table}

\Black{Consequently, the coupled sparse coding problem in Eq.~(\ref{eq:update}) can be compactly reorganized into the following single-variable form:}
\begin{equation}
\label{eq:finalloss}
    \arg\underset{\{\bm{W}\}}{\min} \frac{1}{2} \left\|\bm{Z}-\bm{DW} \right\|_F^2+\lambda\left\|\bm{W}\right\|_1.
\end{equation}

\Black{This reformulation serves as the algorithmic basis for the proposed unfolding architecture. Rather than changing the underlying common/source-specific decomposition prior, it reorganizes the optimization variables so that the corresponding proximal-gradient update can be implemented as a structurally constrained joint unfolding block.}

\subsubsection{Algorithm Unfolding}

We adopt the Learned Iterative Shrinkage-Thresholding Algorithm (LISTA) framework to solve Eq. (\ref{eq:finalloss}). 
Each LISTA iteration consists of a gradient descent step followed by a soft-thresholding operation to enforce $\ell_1$-sparsity. 
The update rule of the combined feature $\bm{W}$ for the $t$-th iteration is:
\begin{equation}
\label{eq:cdupdate}
    \begin{aligned}
    \bm{W}&
    =\mathcal{S}_{\lambda/L}{\left(\bm{W}^{(t-1)}-\frac{1}{L}\bm{D}^{\text{T}}\left(\bm{D}\bm{W}^{(t-1)}-\bm{Z}\right)\right)},
    \end{aligned}
\end{equation}
where $L$ is the Lipschitz constant and $\mathcal{S}_\theta(\cdot)$ denotes the soft-thresholding operator.

The philosophy of deep unfolding is to map the fixed steps of this iterative algorithm into a deep network with learnable parameters. Moreover, 
considering the need for efficient image processing, Learned Convolutional Sparse Coding (LCSC) reformulates the original matrix multiplication-based update into an equivalent convolutional operation:
\begin{equation}
    \label{eq:cdlcscupdate}
    \bm{W}^{(t)}=\mathcal{S}_{\theta^{(t)}}{\left(\bm{W}^{(t-1)}-\bm{d_U}^{(t)}\otimes^{\text{T}}\left(\bm{d_D}^{(t)}\otimes \bm{W}^{(t-1)}-\bm{Z}\right)\right)},
\end{equation}
where $\otimes$ and $\otimes^{\text{T}}$ denote the convolution and transposed convolution operation, $\theta$ is the learnable threshold, $\bm{d_U}^{(t)}, \bm{d_D}^{(t)}$ are the learnable convolutional kernels of the $t$-th layer.

\subsubsection{Explicit Structural Prior}

The convolutional kernels $\bm{d}_U^{(t)}$ and $\bm{d}_D^{(t)}$ are not arbitrary dense kernels. 
\Black{Since the combined dictionary $\bm{D}$ has a block-sparse topology that encodes the common and source-specific decomposition prior, we structurally parameterize $\bm{d}_D^{(t)}$ and $\bm{d}_U^{(t)}$ to mimic the actions of $\bm{D}$ and $\bm{D}^{\text{T}}$ in Eq.~(\ref{eq:cdupdate}):}
\begin{equation}
\label{eq:selfdefine}
\begin{split}
\bm{d_D}^{(t)}\otimes \bm{W} &= \left[\begin{matrix}\bm{d_{UX}}^{(t-\frac{1}{2})} \otimes \bm{Z_X} + \bm{d_{CX}}^{(t-\frac{1}{2})} \otimes \bm{Z_C}\\\bm{d_{UY}}^{(t-\frac{1}{2})} \otimes \bm{Z_Y} + \bm{d_{CY}}^{(t-\frac{1}{2})} \otimes \bm{Z_C}\\ \end{matrix}\right], \\
\bm{d_U}^{(t)}\otimes^{\text{T}} \bm{S} &= \left[\begin{matrix}\bm{d_{UX}}^{(t)} \otimes^{\text{T}} \bm{P} \\ \bm{d_{UY}}^{(t)} \otimes^{\text{T}} \bm{Q} \\ \bm{d_{CX}}^{(t)} \otimes^{\text{T}} \bm{P} + \bm{d_{CY}}^{(t)} \otimes^{\text{T}} \bm{Q} \\ \end{matrix}\right],
\end{split}
\end{equation}
where $\bm{S}=\left[\bm{P},\bm{Q}\right]^{\text{T}}$ is the gradient signal, and $\bm{d_{UX}}$, $\bm{d_{UY}}$, $\bm{d_{CX}}$, and $\bm{d_{CY}}$ are the learnable small convolutional kernels corresponding to the unique and common dictionaries. It should be noted that $\bm{d}_D$ and $\bm{d}_U$ are not required to be exact transpose pairs after end-to-end learning. Instead, they follow the same block-wise interaction topology as $\bm{D}$ and $\bm{D}^{\text{T}}$, which provides a model-derived structural constraint while retaining learnable flexibility.

This design enforces the desired feature decomposition logic while enabling a synchronous, joint update of all components in $\bm{W}$ within a single forward pass. It is the key to CDNet's efficiency, eliminating the sequential computation and parameter overhead of alternating strategies.

\subsection{The Fusion and Reconstruction Module}
\label{sec:FM}

After $T$ CDblocks, we obtain the refined combined feature representation $\bm{W}$. The Fusion and Reconstruction Module generates the final fused feature $\bm{F}$ by reintegrating the original input information via a residual connection:
\begin{equation}
\label{eq:yfuse}
    \bm{F}=\bm{d_{F2}} \otimes \left( \bm{d_{F1}} \otimes \bm{W}+\bm{Z}\right),
\end{equation}
where $\bm{d_{F1}} \in \mathbb{R}^{2C \times 3C\times 1 \times 1 }$ and $\bm{d_{F2}} \in \mathbb{R}^{C \times 2C \times 1 \times 1}$ are point-wise convolution kernels.
The residual connection brings back the original luminance information from $\bm{Z}$ to compensate for the loss during the extraction of the combined feature representation $\bm{W}$. 

\Black{For color image fusion, CDNet performs fusion in the YCbCr space, where the luminance channel is fused by the proposed network, while the chrominance channels are fused using a luminance-consistency-based weighted fusion strategy~\cite{ram2017deepfuse, xu2020u2fusion, zheng2023efficient}.}

\subsection{Loss Functions}
\label{sec:loss}

\begin{figure}[t]
  \centering
  \includegraphics[width=0.4\textwidth]{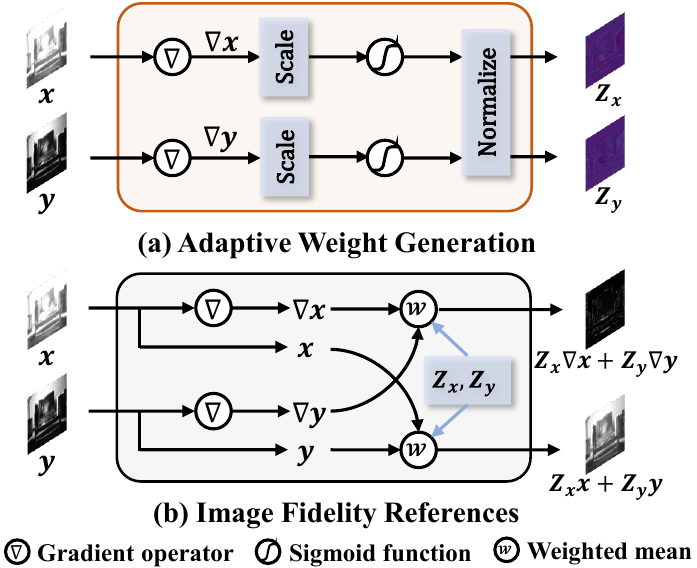}
\caption{Detailed construction of the adaptive references in HLIF. Adaptive weights $\bm{Z}_x$ and $\bm{Z}_y$ are generated from source gradients through scaling, sigmoid mapping, and normalization. The generated weights are then used to construct the high-frequency gradient reference and the low-frequency luminance reference.}

  \label{fig:loss}
\end{figure}

For unsupervised training, we propose a High- and Low-frequency Image Fidelity (HLIF) loss. As detailed in Fig.~\ref{fig:loss}, HLIF first generates adaptive weights from source gradients and then constructs high-frequency gradient and low-frequency luminance references for the fused Y-channel image.

\subsubsection{Adaptive Weight Generation} 
Fusion weights are expected to emphasize salient structural responses while suppressing weak noise-like fluctuations. Given the Y-channel gradient magnitudes $\nabla \bm{x}$ and $\nabla \bm{y}$ computed from the absolute Scharr responses, we observe that raw gradient responses are sensitive to both structural edges and noise. To suppress noise-dominated responses while preserving salient edges, we apply a sigmoid-based gradient-to-weight mapping:
\begin{equation}
\label{eq:weight}
\bm{w_x} = \mathrm{Sigmoid} (\nabla \bm{x}/\tau), \quad \bm{w_y} = \mathrm{Sigmoid} (\nabla \bm{y}/\tau),
\end{equation}
where $\tau$ is a scaling factor. 

It maps gradients to $[0,1]$ with a steep transition zone: low-amplitude noise is dampened in the lower asymptote, while strong edges remain in the active region. Spatial normalization yields adaptive importance maps:
\begin{equation}
\label{eq:weight}
\bm{Z_x} = \frac{\bm{w_x}}{\bm{w_x}+\bm{w_y}}, \quad \bm{Z_y} = \bm{1}-\bm{Z_x}.
\end{equation}

\subsubsection{High-frequency Image Fidelity Loss} 
The High-frequency Image Fidelity (HIF) loss term is designed to maintain consistency in high-frequency components between the fused feature and the input features with a $l_1$-norm loss:
\begin{equation}
\label{eq:hfloss}
\begin{split}
\mathcal{L}_{HIF}&=\left\|\nabla \bm{f}-\left(\bm{Z_x} \odot \nabla\bm{x} + \bm{Z_y} \odot \nabla\bm{y}\right) \right \|_1,
\end{split}
\end{equation}
where $\odot$ denotes the Hadamard product. 
The proposed HIF loss encourages the fused gradient to follow the dominant source, effectively suppressing noise-dominated regions.

\subsubsection{Low-frequency Image Fidelity Loss} 
While $\mathcal{L}_{HIF}$ emphasizes texture structures, it lacks effective constraints on low-frequency components such as overall luminance. To preserve global structure, we introduce a Low-frequency Image Fidelity (LIF) loss operated on image domain:
\begin{equation}
\label{eq:lfloss}
\begin{split}
\mathcal{L}_{LIF}&=\left\|\bm{f}-\left(\bm{Z_x} \odot \bm{x}+\bm{Z_y} \odot \bm{y}\right)\right\|_2,
\end{split}
\end{equation}

In contrast to the $l_1$ norm used in $\mathcal{L}_{HIF}$ for edge preservation, the LIF loss employs the $l_2$ norm. This quadratic penalty effectively suppresses large-area intensity discrepancies, thereby promoting global smoothness and a natural luminance distribution in the low-frequency domain.

\subsubsection{Total Loss} The total High- and Low-frequency Image Fidelity (HLIF) loss is expressed as:
\begin{equation}
\label{eq:totalloss}
\mathcal{L}_{HLIF}=\mathcal{L}_{HIF}+\lambda \mathcal{L}_{LIF},
\end{equation}
where $\lambda$ is a weighting factor setting to 1.0. 
\Black{The HLIF loss provides a simple unsupervised objective by separately constraining gradient-domain structural fidelity and image-domain luminance consistency, without relying on heavy pre-trained networks.}

\begin{table*}[t]
\centering
\renewcommand{\arraystretch}{1.2}
\caption{Datasets used for multi-exposure image fusion (MEF), infrared and visible image fusion (IVF), medical image fusion (MIF), and IVF for segmentation (IVF4Seg) tasks. The notation \enquote{$m$ / $n$} indicates $m$ pairs of images for training and $n$ pairs of images for testing.}
\resizebox{0.85\textwidth}{!}{
\begin{tabular}{l|cc|cc|cc|c}
\toprule[1pt]
 & \multicolumn{2}{c|}{MEF Task} & \multicolumn{2}{c|}{IVF Task} & \multicolumn{2}{c|}{MIF Task} & IVF4Seg Task\\ \midrule 
Datasets & MEFB \cite{zhang2021benchmarking} & SICE \cite{cai2018learning} & TNO \cite{toet2017tno} & RoadScene \cite{xu2020u2fusion} & PET-MRI \cite{harvard} & SPECT-MRI \cite{harvard} & FMB \cite{liu2023multi}\\
Num. of Pairs & 100 & 483 / 118 & 232 & 221 & 269 & 357 & 1220 / 280\\ \bottomrule[1pt]
\end{tabular}
}
\label{tab:dataset} 
\end{table*}

\Black{
\section{Computational Efficiency}
\label{sec:complexity}


In this section, we analyze the computational efficiency of CDNet from an optimization-inspired perspective. 
The purpose of this analysis is not to establish a strict convergence guarantee for the end-to-end trained network, since the convolutional operators in CDBlock are learned and are not constrained to be exact transpose pairs after training. 
Instead, we use the fixed-dictionary sparse coding prototype to interpret the structural origin of CDBlock and then compare the per-stage computational cost of alternating and unified update strategies. 
This analysis characterizes the efficiency benefit brought by the proposed update reorganization, rather than claiming a faster asymptotic convergence rate.

\subsection{Optimization-Bound Interpretation}

Consider the fixed-dictionary sparse coding prototype:
\begin{equation}
F(\bm{W})=
\frac{1}{2}\|\bm{Z}-\bm{D}\bm{W}\|_F^2
+\lambda\|\bm{W}\|_1 ,
\end{equation}
where $\bm{D}$ is the combined dictionary and $\bm{W}$ is the unified representation. 
When $\bm{D}$ is fixed, the smooth term has an $L$-Lipschitz continuous gradient with $L=\|\bm{D}\|_2^2$. 
The standard proximal-gradient update satisfies:
\begin{equation}
F(\bm{W}^{k})-F(\bm{W}^{*})
\leq
\frac{L\|\bm{W}^{0}-\bm{W}^{*}\|_{F}^{2}}{2k},
\quad k\geq 1.
\end{equation}
Therefore, a sufficient number of iterations to guarantee 
$F(\bm{W}^{k})-F(\bm{W}^{*})\leq \epsilon$ is
\begin{equation}
\bar{T}_{\mathrm{Joint}}(\epsilon)
=
\left\lceil
\frac{L\|\bm{W}^{0}-\bm{W}^{*}\|_{F}^{2}}{2\epsilon}
\right\rceil .
\end{equation}
where $\bar{T}_{\mathrm{Joint}}(\epsilon)$ denotes a sufficient worst-case iteration bound, rather than the exact number of iterations required in practice. This bound is used only as an optimization-level reference for the fixed-dictionary prototype. 
After unfolding, the learned convolutional operators relax the exact dictionary-transpose relationship, and therefore the bound should not be interpreted as a formal convergence guarantee for the trained CDNet.

For the alternating strategy, one iteration corresponds to a complete cyclic sweep over the source-specific and common representations. 
Under standard block-wise Lipschitz assumptions, cyclic block proximal-gradient methods also admit an $O(1/k)$ convergence guarantee for convex composite objectives. 
Thus, the joint and alternating strategies have the same sublinear convergence order at the optimization-bound level. 
The advantage of the proposed unified formulation mainly comes from avoiding multiple sequential component updates and reducing the per-stage computational cost, rather than from changing the asymptotic convergence order.

{\subsection{Per-Stage Computational Cost}

We compare the computational cost of the core sparse coding update by counting scalar multiplications and ignoring additions. 
For a multi-modal setting with $N$ input modalities, hidden channel number $C$, spatial size $H\times W$, and kernel size $s$, the alternating strategy sequentially updates $N$ source-specific representations and one common representation within a complete cyclic sweep. 
As derived in the supplementary material, its total multiplication cost is:
\begin{equation}
\mathcal{M}_{\mathrm{AM}}
=
\left(5 + N\right) N s^2 H W C^2 .
\end{equation}

In contrast, the proposed unified strategy updates all source-specific and common representations within a single block-structured sparse coding step, with the cost:
\begin{equation}
\mathcal{M}_{\mathrm{Joint}}
=
4Ns^2HWC^2 .
\end{equation}

Therefore, the relative reduction in scalar multiplications is:
\begin{equation}
\frac{\mathcal{M}_{\mathrm{AM}}-\mathcal{M}_{\mathrm{Joint}}}
{\mathcal{M}_{\mathrm{AM}}}
=
\frac{N+1}{N+5}.
\end{equation}

For the two-source fusion setting considered in this paper, \textit{i.e.}, $N=2$, this corresponds to a theoretical reduction of approximately $42.9\%$ in the core sparse coding update. 
Moreover, since $\frac{N+1}{N+5}$ increases with $N$, the advantage of the unified update becomes more pronounced as the number of input modalities grows. 
This saving comes from the block-sparse dictionary prior, which preserves only the prior-guided block interactions and eliminates the repeated sequential updates required by alternating minimization.

This analysis only characterizes the sparse coding update and should not be interpreted as an exact prediction of the profiled GFLOPs or runtime of the complete network. 
Nevertheless, it explains why the proposed unified update is more compact and more amenable to parallel execution.

\subsection{Practical Efficiency}

For a fair comparison, the alternating-update variant uses the same feature expansion and fusion-reconstruction modules as CDNet, and differs only in the decomposition update strategy. 
Both variants are evaluated under the same input resolution, implementation environment, and inference protocol.
For image pairs of size $256\times256$, the unified update reduces the number of parameters from 7.9K to 6.5K, the average inference time from 2.1097~ms to 1.5508~ms, and the measured computational cost from 1.5~GFLOPs to 0.4~GFLOPs, corresponding to reductions of approximately 18\%, 26\%, and 73\%, respectively. 
These results confirm that the main benefit of the unified update lies in computational efficiency gained from update reorganization, rather than aggressive parameter compression.}

}

\section{Experiments}
\label{sec:experiment}

This section details the experimental setup, evaluates the proposed method against representative state-of-the-art methods across multiple image fusion tasks, and validates its core design choices through ablation studies.

\subsection{Experimental Settings}
\label{sec:setup}

\subsubsection{Datasets} The datasets used for evaluation are summarized in Table \ref{tab:dataset}. \Black{For the SICE dataset, we randomly split the training and testing sets in a 4:1 ratio. 
Since several recent baselines, such as GIFNet and Mask-DiFuser, are limited by memory or implementation constraints when processing the original high-resolution SICE images, we uniformly downsample all SICE test images by a factor of 0.2 before inference. 
This protocol is mainly adopted to ensure that all competing methods can be evaluated under the same memory-feasible resolution. We acknowledge that image resizing may affect high-frequency details, and therefore all methods are evaluated with identical resized inputs for fairness.}

\begin{table}[t]
\centering
\renewcommand{\arraystretch}{1.2}
\caption{The details of baseline methods on MEF, IVF, and MIF tasks. \enquote{Para. (M)} represents the number of parameters in millions, \enquote{Time (ms)} denotes the average inference time.
The best and second-best values are marked in \textbf{bold} and \underline{underline}, respectively.}
\label{tab:baselines}

\resizebox{0.5\textwidth}{!}{
\begin{tabular}{c l c c c}
\toprule[1pt]
\textbf{Task} & \textbf{Method} & \textbf{Venue} & \textbf{Para. (M)$\downarrow$} & \textbf{Time (ms)$\downarrow$} \\
\midrule

\multirow{7}{*}{\rotatebox[origin=c]{90}{\shortstack{\textbf{Multi-exposure} \\ \textbf{Image Fusion}}}}
& HoLoCo \cite{liu2023holoco}       & IF\textquotesingle23     & 17.3874           & 15.0381 \\
& IID-MEF \cite{zhang2023iid}      & IF\textquotesingle23     & 0.3112            & 1233.3202 \\
& FFMEF \cite{zheng2023efficient}  & CVPR\textquotesingle23   & \cellcolor{secondblue}\underline{0.0076}& 14.5022 \\
& DeepM\textsuperscript{2}CDL \cite{deng2023deepm} & TPAMI\textquotesingle24 & 425.1747 & 549.0202 \\
& GIFNet \cite{cheng2025one}       & CVPR\textquotesingle25   & 0.6135            & \cellcolor{secondblue}\underline{13.1098} \\
& Mask-DiFuser \cite{tang2025mask} & TPAMI\textquotesingle25  & 171.2617          & 580.0139 \\
& CDNet                     & Ours      &  \cellcolor{bestblue}\textbf{0.0065}   &  \cellcolor{bestblue}\textbf{1.5508} \\

\midrule

\multirow{7}{*}{\rotatebox[origin=c]{90}{\shortstack{\textbf{Infrared and Visible} \\ \textbf{Image Fusion}}}}
& LRRNet \cite{li2023lrrnet}       & TPAMI\textquotesingle23  & \cellcolor{secondblue}\underline{0.0487}& 18.4793 \\
& CDDFuse \cite{zhao2023cddfuse}   & CVPR\textquotesingle23   & 1.1856            & 25.4836 \\
& EMMA \cite{zhao2024equivariant}  & CVPR\textquotesingle24   & 1.5161            & 25.1692 \\
& DRMF \cite{tang2024drmf}         & MM\textquotesingle24     & 170.8904          & 46.6248 \\
& GIFNet \cite{cheng2025one}       & CVPR\textquotesingle25   & 0.6135            & \cellcolor{secondblue}\underline{13.1098} \\
& Mask-DiFuser \cite{tang2025mask} & TPAMI\textquotesingle25  & 171.2617          & 580.0139 \\
& CDNet                     & Ours      &  \cellcolor{bestblue}\textbf{0.0065}   &  \cellcolor{bestblue}\textbf{1.5508} \\

\midrule

\multirow{7}{*}{\rotatebox[origin=c]{90}{\shortstack{\textbf{Medical} \\ \textbf{Image Fusion}}}}
& LRRNet \cite{li2023lrrnet}       & TPAMI\textquotesingle23  & \cellcolor{secondblue}\underline{0.0487}& 18.4793 \\
& CoCoNet \cite{liu2024coconet}    & IJCV\textquotesingle24   & 0.9114            & \cellcolor{secondblue}\underline{8.7444} \\
& EMMA \cite{zhao2024equivariant}  & CVPR\textquotesingle24   & 1.5161            & 25.1692 \\
& BSAFusion \cite{li2025bsafusion} & AAAI\textquotesingle25   & 9.6942            & 160.9494 \\
& GIFNet \cite{cheng2025one}       & CVPR\textquotesingle25   & 0.6135            & 13.1098 \\
& Mask-DiFuser \cite{tang2025mask} & TPAMI\textquotesingle25  & 171.2617          & 580.0139 \\
& CDNet                     & Ours      &  \cellcolor{bestblue}\textbf{0.0065}   &  \cellcolor{bestblue}\textbf{1.5508} \\

\bottomrule[1pt]
\end{tabular}
}
\end{table}

\begin{table*}[t]
\centering
\renewcommand{\arraystretch}{1.2}
\caption{Quantitative results on the MEF task. The best and second-best values are marked in \textbf{bold} and \underline{underline}, respectively.}
\large 
\resizebox{1\textwidth}{!}{
\begin{tabular}{>{\arraybackslash}p{3.5cm}
>{\centering\arraybackslash}p{1.2cm}
>{\centering\arraybackslash}p{1.2cm}
>{\centering\arraybackslash}p{1.2cm}
>{\centering\arraybackslash}p{1.2cm}
>{\centering\arraybackslash}p{1.2cm}
>{\centering\arraybackslash}p{1.8cm}
>{\centering\arraybackslash}p{1.2cm}
>{\centering\arraybackslash}p{1.2cm}
>{\centering\arraybackslash}p{1.2cm}
>{\centering\arraybackslash}p{1.2cm}
>{\centering\arraybackslash}p{1.2cm}
>{\centering\arraybackslash}p{1.8cm}}
\toprule[1pt]

&\multicolumn{6}{c}{\textbf{MEFB Multi-Exposure Image Fusion Dataset}\cite{zhang2021benchmarking}} & \multicolumn{6}{c}{\textbf{SICE Multi-Exposure Image Fusion Dataset} \cite{cai2018learning}}\\
 & MSE$\downarrow$ & PSNR$\uparrow$ & SSIM$\uparrow$ & CC$\uparrow$ & Nabf$\downarrow$ & HyperIQA$\uparrow$ & MSE$\downarrow$ & PSNR$\uparrow$ & SSIM$\uparrow$ & CC$\uparrow$ & Nabf$\downarrow$ & HyperIQA$\uparrow$\\ \midrule \midrule
HoLoCo \cite{liu2023holoco} & 0.10 & 58.44 & 0.34 & 0.89 & 0.04 & 50.39 & 0.12 & 57.80 & 0.24 & 0.86 & 0.01 & 54.04\\
IID-MEF \cite{zhang2023iid} & \cellcolor{secondblue}\underline{0.09} & \cellcolor{secondblue}\underline{58.95} & 0.39 & 0.90 & 0.01 & 55.50 & 0.15 & 56.99 & 0.29 & 0.87 & \cellcolor{secondblue}\underline{0.01} & 52.66\\
FFMEF \cite{zheng2023efficient} & 0.10 & 58.72 &  \cellcolor{bestblue}\textbf{0.46} & \cellcolor{secondblue}\underline{0.90} & \cellcolor{secondblue}\underline{0.01} & 56.59 & \cellcolor{secondblue}\underline{0.11} & \cellcolor{secondblue}\underline{58.26} & 0.30 & \cellcolor{secondblue}\underline{0.89} & 0.01 & 58.47\\
DeeM\textsuperscript{2}CDL \cite{deng2023deepm} & 0.18 & 56.01 & 0.40 & 0.85 & 0.06 & 46.69 & 0.20 & 55.46 & 0.32 & 0.83 & 0.05 & 51.87\\
GIFNet \cite{cheng2025one} & 0.11 & 58.15 & 0.31 & 0.89 & 0.12 & \cellcolor{secondblue}\underline{56.64} & 0.14 & 57.38 & 0.22 & 0.84 & 0.13 & \cellcolor{secondblue}\underline{59.65}\\
Mask-DiFuser \cite{tang2025mask} & 0.11 & 58.04 & 0.35 & 0.89 & 0.06 & 54.78 & 0.13 & 57.22 & \cellcolor{secondblue}\underline{0.36} & 0.88 & 0.05 & 58.27\\
CDNet (Ours) &  \cellcolor{bestblue}\textbf{0.09} &  \cellcolor{bestblue}\textbf{59.06} & \cellcolor{secondblue}\underline{0.45} &  \cellcolor{bestblue}\textbf{0.91} &  \cellcolor{bestblue}\textbf{0.01} &  \cellcolor{bestblue}\textbf{58.03} &  \cellcolor{bestblue}\textbf{0.11} &  \cellcolor{bestblue}\textbf{58.28} &  \cellcolor{bestblue}\textbf{0.43} &  \cellcolor{bestblue}\textbf{0.89} &  \cellcolor{bestblue}\textbf{0.01} &  \cellcolor{bestblue}\textbf{60.89}\\ \bottomrule[1pt]
\end{tabular}
}
\label{tab:mefres}
\end{table*}

\subsubsection{Implementation Details} 
The decomposition stage of CDNet consists of $n=3$ CDBlocks, with dictionary kernel size $s=3$ and channel number $C=5$. 
All experiments are conducted on a computer with an NVIDIA GeForce RTX 4090 GPU.
The Adam optimizer is used for training. The batch size and the learning rate are set to 10 and 0.0005, respectively. The weighted fusion formula for the chrominance channels adopts the implementation from~\cite{ram2017deepfuse, xu2020u2fusion, zheng2023efficient}. 
Under an all-in-one setting, CDNet is trained for 50 epochs only on the SICE training set, and evaluated without fine-tuning on heterogeneous tasks.
The hyperparameter tuning process of CDNet is presented in the supplementary materials.

\subsubsection{Evaluation Metrics} \Black{We employ commonly used fusion metrics, including MSE, PSNR, SSIM, CC, Nabf, and HyperIQA \cite{jagalingam2015review, shreyamsha2013multifocus, su2020blindly}. 
Since ground-truth fused images are generally unavailable in multi-source image fusion, MSE, PSNR, and SSIM are computed by averaging the scores between the fused image and each source image, following common source-referenced evaluation protocols in image fusion \cite{li2023lrrnet, wang2024new, liu2024coconet, li2025bsafusion}. 
Similarly, CC and Nabf are used to measure source information preservation and fusion artifacts, while HyperIQA provides a no-reference perceptual quality assessment. We note that source-referenced metrics cannot fully characterize perceptual fusion quality; therefore, we also report no-reference perceptual assessment and downstream segmentation performance as complementary evaluations.}

\subsubsection{Comparison Methods} 
\Black{For competing methods, we use the official checkpoints released by the original authors whenever available, and keep their original inference procedures unchanged.} We compare CDNet with other methods on multiple tasks. For the multi-exposure image fusion (MEF) task, the comparison methods include HoLoCo \cite{liu2023holoco}, IID-MEF \cite{zhang2023iid}, FFMEF \cite{zheng2023efficient}, DeepM\textsuperscript{2}CDL \cite{deng2023deepm}, GIFNet \cite{cheng2025one}, and Mask-DiFuser \cite{tang2025mask}. For the infrared and visible image fusion (IVF) task, the comparison methods include LRRNet \cite{li2023lrrnet}, CDDFuse \cite{zhao2023cddfuse}, EMMA \cite{zhao2024equivariant}, DRMF \cite{tang2024drmf}, GIFNet \cite{cheng2025one}, and Mask-DiFuser \cite{tang2025mask}. For the medical image fusion (MIF) task, the comparison methods include LRRNet \cite{li2023lrrnet}, CoCoNet \cite{liu2024coconet}, EMMA \cite{zhao2024equivariant}, BSAFusion \cite{li2025bsafusion}, GIFNet \cite{cheng2025one}, and Mask-DiFuser \cite{tang2025mask}.

\begin{figure*}[t]
  \centering
  \includegraphics[width=0.88\textwidth]{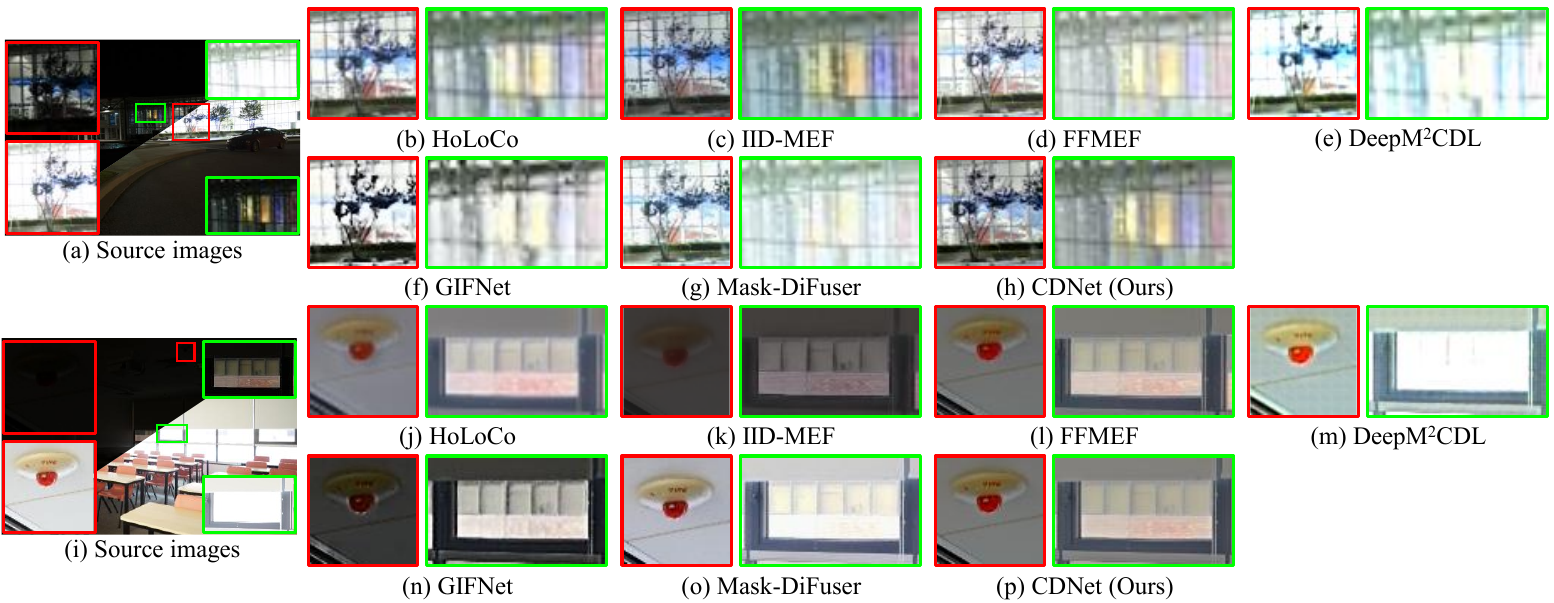}
  \caption{Visual comparison for \enquote{Zentrum} in MEFB dataset and \enquote{SICE-Dataset\_Part1\_12} in SICE dataset.}
  \label{fig:Qual_MEF}
\end{figure*}

\begin{table*}[t]
\centering
\renewcommand{\arraystretch}{1.2}
\caption{Quantitative results on the IVF task. The best and second-best values are marked in \textbf{bold} and \underline{underline}, respectively.}
\large 
\resizebox{1\textwidth}{!}{
\begin{tabular}{>{\arraybackslash}p{3.5cm}
>{\centering\arraybackslash}p{1.2cm}
>{\centering\arraybackslash}p{1.2cm}
>{\centering\arraybackslash}p{1.2cm}
>{\centering\arraybackslash}p{1.2cm}
>{\centering\arraybackslash}p{1.2cm}
>{\centering\arraybackslash}p{1.8cm}
>{\centering\arraybackslash}p{1.2cm}
>{\centering\arraybackslash}p{1.2cm}
>{\centering\arraybackslash}p{1.2cm}
>{\centering\arraybackslash}p{1.2cm}
>{\centering\arraybackslash}p{1.2cm}
>{\centering\arraybackslash}p{1.8cm}}
\toprule[1pt]
& \multicolumn{6}{c}{\textbf{TNO Infrared-Visible Image Fusion Dataset} \cite{toet2017tno}} & \multicolumn{6}{c}{\textbf{RoadScene Infrared-Visible Image Fusion Dataset} \cite{xu2020u2fusion}}\\
& MSE$\downarrow$ & PSNR$\uparrow$ & SSIM$\uparrow$ & CC$\uparrow$ & Nabf$\downarrow$ & HyperIQA$\uparrow$ & MSE$\downarrow$ & PSNR$\uparrow$ & SSIM$\uparrow$ & CC$\uparrow$ & Nabf$\downarrow$ & HyperIQA$\uparrow$\\ \midrule \midrule
LRRNet \cite{li2023lrrnet} & 0.06 & 61.29 & 0.32 & 0.43 & 0.10 & 38.12 & 0.07 & 59.93 & 0.33 & 0.62 & \cellcolor{secondblue}\underline{0.04} & 34.54\\
CDDFuse \cite{zhao2023cddfuse} & 0.05 & 61.92 & \cellcolor{secondblue}\underline{0.48} & 0.47 & 0.09 & 37.26 & 0.05 & 62.12 & \cellcolor{secondblue}\underline{0.48} & 0.63 & 0.09 & 37.59\\
EMMA \cite{zhao2024equivariant} & 0.05 & 61.84 & 0.42 & 0.46 & 0.10 & 36.63 & 0.05 & 62.06 & 0.46 & 0.62 & 0.08 & 35.05\\
DRMF \cite{tang2024drmf} & 0.07 & 60.14 & 0.29 & 0.36 & 0.10 & 36.50 & 0.07 & 60.55 & 0.42 & 0.50 & 0.06 & 35.20\\
GIFNet \cite{cheng2025one} & \cellcolor{secondblue}\underline{0.03} & \cellcolor{secondblue}\underline{64.04} & 0.44 & 0.49 & \cellcolor{secondblue}\underline{0.07} &  \cellcolor{bestblue}\textbf{42.33} & \cellcolor{secondblue}\underline{0.04} & \cellcolor{secondblue}\underline{62.85} & 0.43 & 0.64 & 0.10 &  \cellcolor{bestblue}\textbf{39.23}\\
Mask-DiFuser \cite{tang2025mask} & 0.05 & 61.63 & 0.36 &  \cellcolor{bestblue}\textbf{0.53} & 0.23 & 37.82 & 0.05 & 61.05 & 0.44 & \cellcolor{secondblue}\underline{0.64} & 0.09 & 35.08\\
CDNet (Ours) &  \cellcolor{bestblue}\textbf{0.02} &  \cellcolor{bestblue}\textbf{65.27} &  \cellcolor{bestblue}\textbf{0.54} & \cellcolor{secondblue}\underline{0.51} &  \cellcolor{bestblue}\textbf{0.00}$_{\textbf{12}}$ & \cellcolor{secondblue}\underline{41.07} &  \cellcolor{bestblue}\textbf{0.03} &  \cellcolor{bestblue}\textbf{64.44} &  \cellcolor{bestblue}\textbf{0.51} &  \cellcolor{bestblue}\textbf{0.66} &  \cellcolor{bestblue}\textbf{0.00}$_{\textbf{46}}$ & \cellcolor{secondblue}\underline{38.32}\\ \bottomrule
\end{tabular}
}
\label{tab:ivfres}
\end{table*}

\begin{figure*}[t]
  \centering
\includegraphics[width=0.88\textwidth]{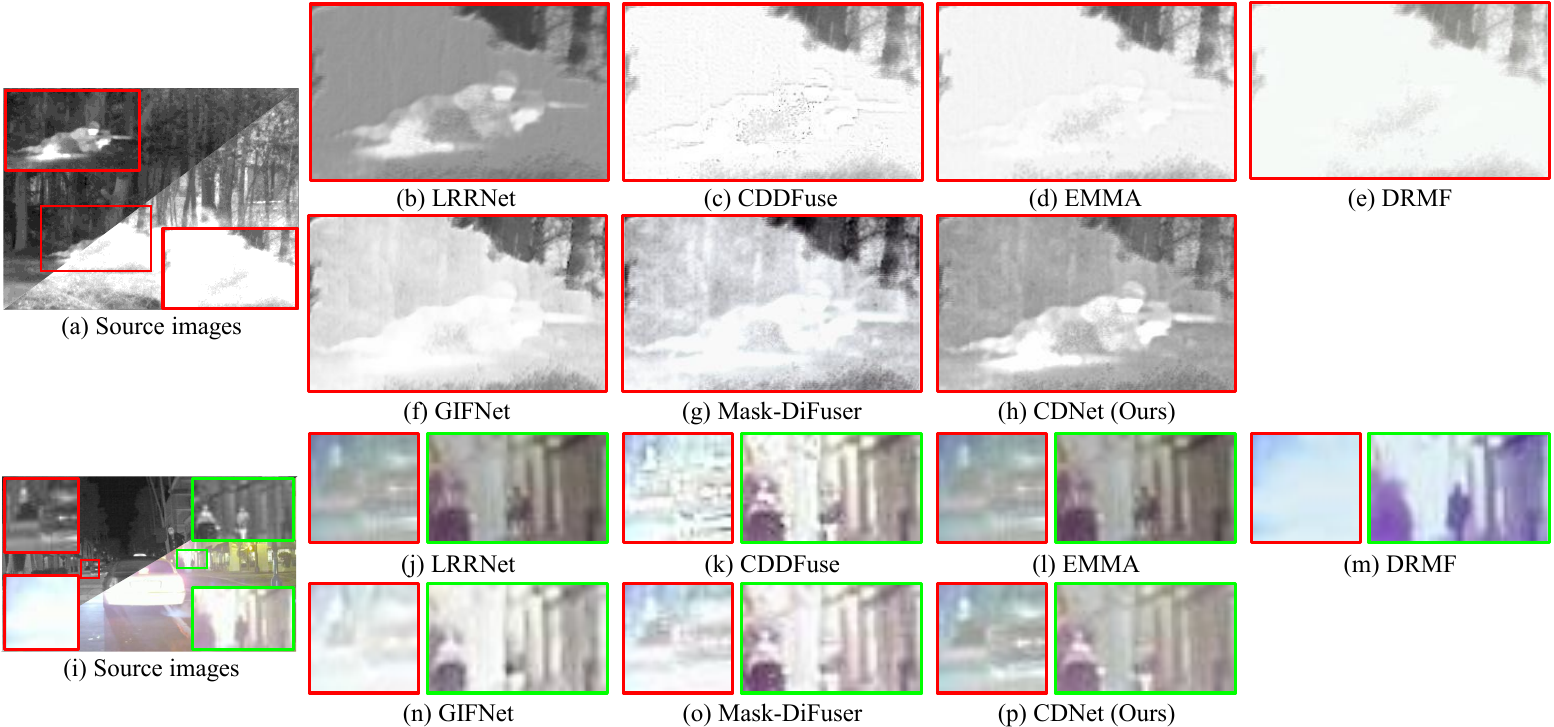}
  \caption{Visual comparison for \enquote{soldier\_behind\_smoke\_1} in TNO dataset and \enquote{FLIR\_01415} in RoadScene dataset.}
  \label{fig:Qual_IVF}
\end{figure*}

\begin{table*}[t]
\centering
\renewcommand{\arraystretch}{1.2}
\caption{Quantitative results on the MIF task. The best and second-best values are marked in \textbf{bold} and \underline{underline}, respectively.
}
\large 
\resizebox{1\textwidth}{!}{
\begin{tabular}{>{\arraybackslash}p{3.5cm}
>{\centering\arraybackslash}p{1.2cm}
>{\centering\arraybackslash}p{1.2cm}
>{\centering\arraybackslash}p{1.2cm}
>{\centering\arraybackslash}p{1.2cm}
>{\centering\arraybackslash}p{1.2cm}
>{\centering\arraybackslash}p{1.8cm}
>{\centering\arraybackslash}p{1.2cm}
>{\centering\arraybackslash}p{1.2cm}
>{\centering\arraybackslash}p{1.2cm}
>{\centering\arraybackslash}p{1.2cm}
>{\centering\arraybackslash}p{1.2cm}
>{\centering\arraybackslash}p{1.8cm}}
\toprule[1pt]
& \multicolumn{6}{c}{\textbf{PET-MRI Medical Image Fusion Dataset} \cite{harvard}} & \multicolumn{6}{c}{\textbf{SPECT-MRI Medical Image Fusion Dataset} \cite{harvard}}\\
& MSE$\downarrow$ & PSNR$\uparrow$ & SSIM$\uparrow$ & CC$\uparrow$ & Nabf$\downarrow$ & HyperIQA$\uparrow$ & MSE$\downarrow$ & PSNR$\uparrow$ & SSIM$\uparrow$ & CC$\uparrow$ & Nabf$\downarrow$ & HyperIQA$\uparrow$\\ \midrule \midrule
LRRNet \cite{li2023lrrnet} & 0.05 & 61.56 & 0.30 & 0.80 & 0.01 & 35.66 & 0.02 & 64.59 & 0.13 & 0.86 & 0.03 & 35.24\\
CoCoNet \cite{liu2024coconet} & 0.08 & 59.40 & 0.24 & 0.82 & 0.01 & 28.55 & 0.09 & 58.62 & 0.18 & 0.89 & 0.05 & 34.45\\
EMMA \cite{zhao2024equivariant} & 0.06 & 60.78 & 0.41 & 0.80 & 0.01 & 36.96 & 0.02 & 65.96 & 0.38 & 0.88 & 0.03 & 34.40\\
BSAFusion \cite{li2025bsafusion} & 0.06 & 60.53 &  \cellcolor{bestblue}\textbf{0.60} & 0.78 & 0.01 & \cellcolor{secondblue}\underline{37.33} & 0.02 & 66.22 & \cellcolor{secondblue}\underline{0.58} & 0.85 & 0.04 & 35.51\\
GIFNet \cite{cheng2025one} & 0.04 & 62.09 & 0.25 & 0.81 & 0.04 &  \cellcolor{bestblue}\textbf{39.59} & \cellcolor{secondblue}\underline{0.01} & \cellcolor{secondblue}\underline{67.69} & 0.23 & 0.88 & 0.05 &  \cellcolor{bestblue}\textbf{39.34}\\
Mask-DiFuser \cite{tang2025mask} & \cellcolor{secondblue}\underline{0.03} & \cellcolor{secondblue}\underline{62.80} & 0.32 & \cellcolor{secondblue}\underline{0.83} & \cellcolor{secondblue}\underline{0.01} & 36.45 & 0.01 & 66.86 & 0.28 & \cellcolor{secondblue}\underline{0.90} & \cellcolor{secondblue}\underline{0.01} & 37.04\\
CDNet (Ours) &  \cellcolor{bestblue}\textbf{0.03} &  \cellcolor{bestblue}\textbf{63.44} & \cellcolor{secondblue}\underline{0.57} &  \cellcolor{bestblue}\textbf{0.84} &  \cellcolor{bestblue}\textbf{0.00}$_{\textbf{14}}$ & 35.01 &  \cellcolor{bestblue}\textbf{0.01} &  \cellcolor{bestblue}\textbf{69.35} &  \cellcolor{bestblue}\textbf{0.59} &  \cellcolor{bestblue}\textbf{0.91} &  \cellcolor{bestblue}\textbf{0.00}$_{\textbf{14}}$ & \cellcolor{secondblue}\underline{37.19}\\ \bottomrule[1pt]
\end{tabular}
}
\label{tab:mifres}
\end{table*}

\begin{figure*}[t]
  \centering
\includegraphics[width=0.88\textwidth]{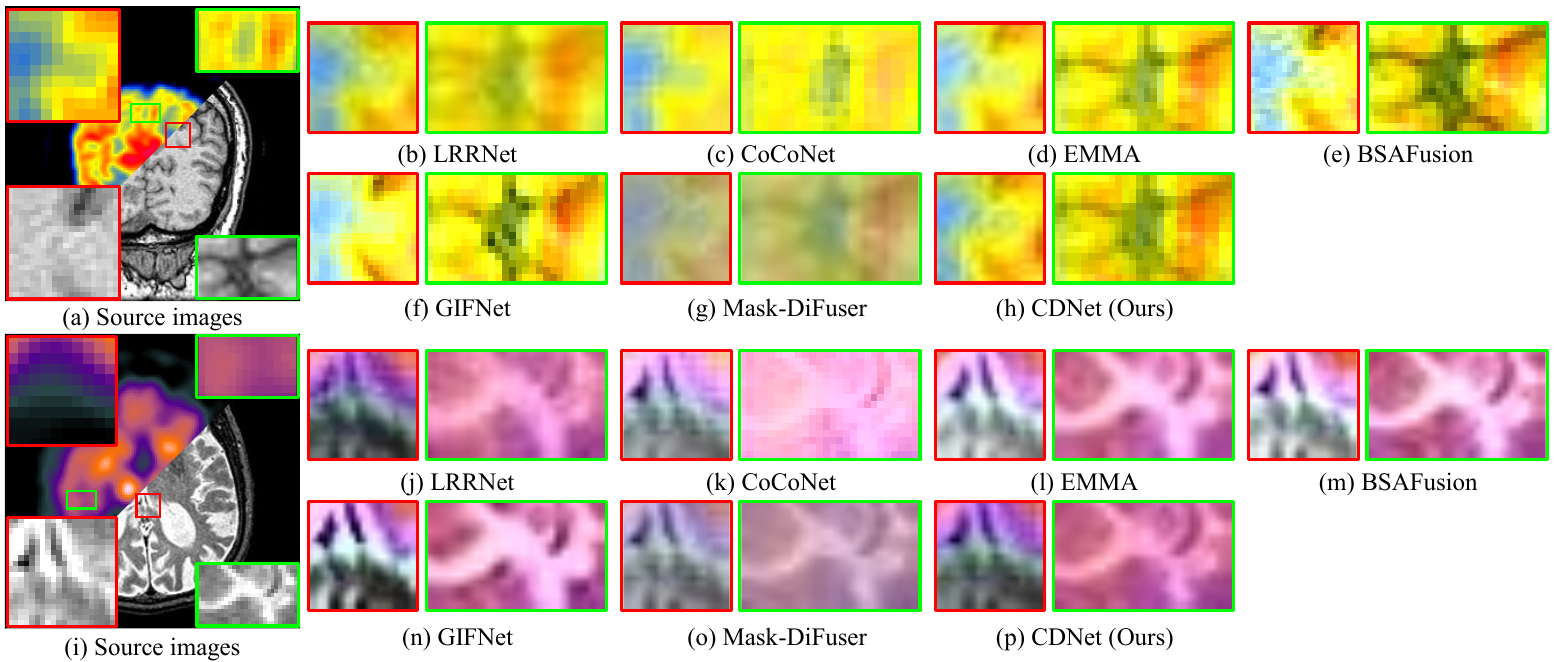}
  \caption{Visual comparison for \enquote{25026} in PET-MRI dataset and \enquote{3025} in SPECT-MRI dataset.}
  \label{fig:Qual_MIF}
\end{figure*}

\subsection{Efficiency Analysis}

The extreme lightweight design of CDNet is its defining characteristic. As summarized in Table \ref{tab:baselines}, CDNet contains around 6.5K parameters. Consequently, it achieves the fastest inference time (1.55 ms for 256$\times$256 images). This exceptional compactness and speed stem directly from the unified update framework and processing of only the Y channel, validating the core efficiency objective of our design.

\subsection{Results on Multi-Exposure Image Fusion Task}
\label{sec:MEF_expe}

\subsubsection{Quantitative Comparison} The quantitative comparison results of all methods are presented in Table~\ref{tab:mefres}. On the MEFB dataset, CDNet achieves the best scores on five metrics: MSE, PSNR, CC, Nabf, and HyperIQA, and obtains the second-best score on SSIM. Notably, it outperforms the second-best method by 1.39 on HyperIQA, indicating that the images fused by CDNet align more closely with human subjective perception. On the SICE dataset, CDNet achieves the best performance on all six metrics. 
These results indicate that the fused images obtained by CDNet contain more details, less noise, and appear more natural.

\subsubsection{Qualitative Comparison} We qualitatively compare the performance of various methods on the MEF task, with results shown in Fig.~\ref{fig:Qual_MEF}. In MEFB dataset, the images fused by HoLoCo \cite{liu2023holoco}, FFMEF \cite{zheng2023efficient}, DeepM\textsuperscript{2}CDL \cite{deng2023deepm}, GIFNet \cite{cheng2025one}, and Mask-DiFuser \cite{tang2025mask} exhibit noticeable exposure inconsistencies in the tree and window regions (highlighted in the red and green boxes), indicating insufficient detail preservation. IID-MEF \cite{zhang2023iid} preserves textures more effectively, yet their results appear blurrier compared to CDNet. In SICE dataset, the results of IID-MEF \cite{zhang2023iid}, DeepM\textsuperscript{2}CDL \cite{deng2023deepm}, GIFNet \cite{cheng2025one}, and Mask-DiFuser \cite{tang2025mask} suffer from exposure inconsistencies in the warning light and window regions. Moreover, the images generated by HoLoCo \cite{liu2023holoco}, FFMEF \cite{zheng2023efficient} show blurred texture details, while CDNet produces clearer details. Overall, CDNet demonstrates superior ability in fusing well-exposed regions from the source images.

\subsection{Results on Infrared and Visible Image Fusion Task}
\label{sec:IVF_expe}

\subsubsection{Quantitative Comparison} The quantitative comparison results of all methods are shown in Table~\ref{tab:ivfres}. On the TNO dataset, CDNet achieves the best performance on four metrics: MSE, PSNR, SSIM, and Nabf, and ranks second on the remaining two metrics, CC and HyperIQA. Notably, CDNet surpasses the second-best method by 1.23 dB in PSNR, and its Nabf value is as low as 0.0012, close to zero, indicating that the fused images contain minimal artifacts or structural distortions. On the RoadScene dataset, CDNet obtains the highest scores on five metrics: MSE, PSNR, SSIM, CC, and Nabf, and achieves the second-best result on HyperIQA. In particular, it outperforms the runner-up by 1.59 dB in PSNR, with an extremely low Nabf of 0.0046, further demonstrating its superior fidelity and robustness in preserving source image content.
Although CDNet is trained on the SICE dataset, it performs effectively on the IVF task, demonstrating strong capability in fusing texture details from source images and exhibiting excellent robustness and transferability.

\subsubsection{Qualitative Comparison} We qualitatively compare the performance of various methods on the IVF task, with results shown in Fig.~\ref{fig:Qual_IVF}. In TNO dataset, the images fused by CDDFuse \cite{zhao2023cddfuse}, EMMA \cite{zhao2024equivariant}, DRMF \cite{tang2024drmf}, GIFNet \cite{cheng2025one}, and Mask-DiFuser \cite{tang2025mask} exhibit poor fusion quality (highlighted in the red box), where soldiers behind smoke are nearly invisible. LRRNet \cite{li2023lrrnet} preserves the texture of the soldier more effectively, but still fails to mitigate the occlusion caused by the smoke in surrounding regions. In contrast, CDNet achieves the best texture preservation, clearly revealing both the soldier and background details. In RoadScene dataset, CDDFuse \cite{zhao2023cddfuse}, DRMF \cite{tang2024drmf}, GIFNet \cite{cheng2025one}, and Mask-DiFuser \cite{tang2025mask} fail to adequately preserve the textures of vehicles and buildings in over-exposed regions (highlighted in the red and green boxes), losing critical information from the infrared image. While LRRNet \cite{li2023lrrnet} and EMMA \cite{zhao2024equivariant} retain more details, their results remain less sharp compared to CDNet. Overall, CDNet demonstrates superior detail and texture preservation across challenging scenarios.

\subsection{Results on Medical Image Fusion Task}
\label{sec:MIF_expe}

\subsubsection{Quantitative Comparison} The quantitative comparison results of all methods are shown in Table~\ref{tab:mifres}. On the PET-MRI dataset, CDNet achieves the best performance on four metrics: MSE, PSNR, CC, and Nabf, and ranks second on SSIM. Notably, it outperforms the second-best method by 0.64 dB in PSNR. On the SPECT-MRI dataset, CDNet obtains the highest scores on five metrics: MSE, PSNR, SSIM, CC, and Nabf, and achieves the second-best result on HyperIQA. In particular, it surpasses the runner-up by 1.66 dB in PSNR. The improvement in PSNR across both datasets indicates that CDNet introduces minimal noise and preserves high fidelity during the fusion process.

\subsubsection{Qualitative Comparison} Fig.~\ref{fig:Qual_MIF} presents the qualitative performance of various methods on the MIF task. In PET-MRI and SPECT-MRI datasets, the images fused by LRRNet \cite{li2023lrrnet}, CoCoNet \cite{liu2024coconet}, and Mask-DiFuser \cite{tang2025mask} exhibit blurred brain structures (highlighted in the red and green boxes), indicating poor texture preservation. EMMA \cite{zhao2024equivariant}, BSAFusion \cite{li2025bsafusion}, and GIFNet \cite{cheng2025one} preserve textures more effectively, yet their results appear slightly less sharp than those of CDNet. In contrast, the image fused by CDNet appears more natural and preserves sharper details, facilitating better visualization and potential lesion localization.

\begin{table*}[]
\centering
\Huge
\renewcommand{\arraystretch}{1.2}
\caption{Quantitative results (\%) on the IVF for semantic segmentation task. The best and second-best values are marked in \textbf{bold} and \underline{underline}, respectively.}
\resizebox{1\textwidth}{!}{
\begin{tabular}{lc@{\hspace{0.15em}}cc@{\hspace{0.15em}}cc@{\hspace{0.15em}}cc@{\hspace{0.15em}}cc@{\hspace{0.15em}}cc@{\hspace{0.15em}}cc@{\hspace{0.15em}}cc@{\hspace{0.15em}}cc@{\hspace{0.15em}}c|cc}
\toprule[1pt]
 & \multicolumn{2}{c}{Road} & \multicolumn{2}{c}{Sidewalk} & \multicolumn{2}{c}{Building} & \multicolumn{2}{c}{T-Lamp} & \multicolumn{2}{c}{Vegetation} & \multicolumn{2}{c}{Person} & \multicolumn{2}{c}{Car} & \multicolumn{2}{c}{Bus} & \multicolumn{2}{c|}{Motorcycle} &  &  \\
\multirow{-2}{*}{Methods} & IoU$\uparrow$ & Acc$\uparrow$ & IoU$\uparrow$ & Acc$\uparrow$ & IoU$\uparrow$ & Acc$\uparrow$ & IoU$\uparrow$ & Acc$\uparrow$ & IoU$\uparrow$ & Acc$\uparrow$ & IoU$\uparrow$ & Acc$\uparrow$ & IoU$\uparrow$ & Acc$\uparrow$ & IoU$\uparrow$ & Acc$\uparrow$ & IoU$\uparrow$ & Acc$\uparrow$ & \multirow{-2}{*}{mIoU$\uparrow$} & \multirow{-2}{*}{mAcc$\uparrow$} \\ \midrule \midrule
LRRNet \cite{li2023lrrnet} & 12.1 & 12.8 & 79.1 & 97.6 & 79.1 & 89.4 & 4.7 & 4.8 & 79.7 & 89.8 & 44.1 & 50.4 & 77.2 & 91.8 & \cellcolor{secondblue}\underline{58.3} & \cellcolor{secondblue}\underline{68.8} & 37.3 & 41.4 & 50.6 & 58.0 \\
CDDFuse \cite{zhao2023cddfuse} & 10.8 & 11.7 & 76.9 & \cellcolor{secondblue}\underline{97.7} & 78.6 & 88.1 & 5.2 & 5.3 & 80.9 & 89.6 & 33.0 & 35.9 & 75.6 & 90.8 & 48.8 & 57.0 & 37.2 & 41.5 & 48.6 & 55.8 \\
EMMA \cite{zhao2024equivariant} & 12.2 & 13.1 & 78.8 & 97.5 & 78.8 & 89.0 & 5.2 & 5.2 & 80.9 & 89.9 & 39.3 & 43.9 & 77.0 & 91.4 & 54.5 & 63.2 & 35.6 & 38.7 & 50.3 & 57.4 \\
DRMF \cite{tang2024drmf} & \cellcolor{secondblue}\underline{14.2} &  \cellcolor{bestblue}\textbf{17.8} &  \cellcolor{bestblue}\textbf{85.0} & 97.2 & \cellcolor{secondblue}\underline{80.8} & 89.6 & 21.3 & 22.0 & \cellcolor{secondblue}\underline{85.1} & 92.0 & 43.2 & 47.9 & \cellcolor{secondblue}\underline{77.3} & \cellcolor{secondblue}\underline{92.8} & 45.6 & 52.1 &  \cellcolor{bestblue}\textbf{46.9} &  \cellcolor{bestblue}\textbf{59.4} & \cellcolor{secondblue}\underline{56.0} & \cellcolor{secondblue}\underline{63.9} \\
GIFNet \cite{cheng2025one} & 13.0 & \cellcolor{secondblue}\underline{17.3} & 82.7 & 97.0 & 77.6 &  \cellcolor{bestblue}\textbf{91.3} & 24.5 & 25.6 & 74.5 & 93.0 & \cellcolor{secondblue}\underline{47.5} & \cellcolor{secondblue}\underline{53.4} & 76.1 &  \cellcolor{bestblue}\textbf{93.4} & 56.4 & 61.9 & \cellcolor{secondblue}\underline{44.5} & 57.1 & 53.6 & 62.7 \\
Mask-DiFuser \cite{tang2025mask} & 13.2 & 16.5 & 81.3 & 97.5 & 79.5 & 90.7 &  \cellcolor{bestblue}\textbf{28.2} &  \cellcolor{bestblue}\textbf{30.1} & 84.5 & \cellcolor{secondblue}\underline{93.0} & 40.7 & 44.5 & 75.1 & 91.1 & 51.4 & 56.1 & 41.1 & 54.8 & 54.3 & 62.4 \\
CDNet (Ours) &  \cellcolor{bestblue}\textbf{14.9} & 17.2 & \cellcolor{secondblue}\underline{82.9} &  \cellcolor{bestblue}\textbf{97.8} &  \cellcolor{bestblue}\textbf{81.5} & \cellcolor{secondblue}\underline{91.1} & \cellcolor{secondblue}\underline{27.0} & \cellcolor{secondblue}\underline{28.3} &  \cellcolor{bestblue}\textbf{85.7} &  \cellcolor{bestblue}\textbf{93.6} &  \cellcolor{bestblue}\textbf{49.9} &  \cellcolor{bestblue}\textbf{55.0} &  \cellcolor{bestblue}\textbf{78.0} & 91.6 &  \cellcolor{bestblue}\textbf{61.9} &  \cellcolor{bestblue}\textbf{70.8} & 43.0 & \cellcolor{secondblue}\underline{58.2} &  \cellcolor{bestblue}\textbf{57.0} &  \cellcolor{bestblue}\textbf{65.1}\\
\bottomrule[1pt]

\end{tabular}
}
\label{tab:seg}
\end{table*}

\subsection{Results on Infrared and Visible Image Fusion for Segmentation Task}
\label{sec:extension}

\Black{To further evaluate the quality of the fused images obtained by different image fusion methods}, we conducted semantic segmentation on the FMB dataset evaluated with respect to both Acc and IoU metrics. The segmentation model used is SegFormer~\cite{xie2021segformer}, and we retrained it on the FMB training set.
The experimental results (\%) are shown in Table \ref{tab:seg}. 
\Black{Our method achieves superior performance in mIoU and mAcc, exceeding the second-best method by 1.0\% and 1.2\%, respectively}. Notably, it leads in several key categories (\textit{e.g.}, Building, Vegetation, Person, Bus). These results further indicate that CDNet can provide fused images beneficial to downstream semantic segmentation.

\subsection{Ablation Studies}
\label{sec:ablation}

In this section, we conduct ablation experiments mainly on the SICE test set to validate the effectiveness and robustness of CDNet, the transferability of HLIF, and the transferability of their joint design under different training datasets.
\Black{More ablation studies, including those on loss terms and the sigmoid operator, are provided in the supplementary materials.}

\begin{table}[]
\centering
\renewcommand{\arraystretch}{1.2}
\caption{Ablation results of update strategies evaluated on the SICE test set.}
\Large
\resizebox{0.45\textwidth}{!}{
\begin{tabular}{lccccc}
\toprule[1pt]
Config. & Para. (K)$\downarrow$ & Time (ms)$\downarrow$ & GFLOPs$\downarrow$& PSNR$\uparrow$ & HyperIQA$\uparrow$ \\ \midrule \midrule
Alter. & 7.9 & 2.1097 & 1.5 & 58.12 & 60.04 \\
Unified &  \cellcolor{bestblue}\textbf{6.5} &  \cellcolor{bestblue}\textbf{1.5508} &  \cellcolor{bestblue}\textbf{0.4} &  \cellcolor{bestblue}\textbf{58.28} &  \cellcolor{bestblue}\textbf{60.89} \\ \bottomrule[1pt]
\end{tabular}
}

\label{tab:ablation1}
\end{table}

\subsubsection{Update Strategies} 
The update strategy influences both performance and efficiency. As shown in Table~\ref{tab:ablation1}, the unified update achieves a higher PSNR of 58.28 dB compared to 58.12 dB for the alternating approach, and yields a notably better HyperIQA score of 60.89 versus 60.04, indicating improved reconstruction fidelity and perceptual quality. 

From a model complexity perspective, the unified scheme employs fewer parameters (6.5K \text{vs.} 7.9K), achieving an approximately 18\% reduction in model size. When fusing a pair of $256 \times 256$ images, this compact formulation translates into substantially improved computational efficiency. Specifically, the average inference time is further decreased by about 26\% (1.5508~ms \text{vs.} 2.1097~ms), and the computational cost is reduced from 1.5~GFLOPs to 0.4~GFLOPs (approximately 73\%). This efficiency gain arises from avoiding the repeated access and sequential update of multiple intermediate feature representations inherent in the alternating strategy, which limits parallelism and incurs additional memory access overhead.

\begin{table}[]
\centering
\renewcommand{\arraystretch}{1.2}
\caption{Ablation results of loss functions on the SICE test set.}
\Large
\resizebox{0.45\textwidth}{!}{
\begin{tabular}{lcccccc}
\toprule[1pt]

Config. & MSE$\downarrow$ & PSNR$\uparrow$ & SSIM$\uparrow$ & CC$\uparrow$ & Nabf$\downarrow$ & HyperIQA$\uparrow$ \\ \midrule \midrule
SSIM-MSE & 0.12 & 58.18 &  \cellcolor{secondblue}\underline{0.42} & 0.86 &  \cellcolor{secondblue}\underline{0.01} & 57.09 \\
VGG & 0.11 & 58.20 & 0.33 & 0.85 & 0.03 &  \cellcolor{secondblue}\underline{59.23} \\
GIF &  \cellcolor{secondblue}\underline{0.11} &  \cellcolor{secondblue}\underline{58.25} & 0.41 &  \cellcolor{secondblue}\underline{0.87} & 0.04 & 58.75 \\
HLIF &  \cellcolor{bestblue}\textbf{0.11} &  \cellcolor{bestblue}\textbf{58.28} &  \cellcolor{bestblue}\textbf{0.43} &  \cellcolor{bestblue}\textbf{0.89} &  \cellcolor{bestblue}\textbf{0.01} &  \cellcolor{bestblue}\textbf{60.89} \\ \bottomrule[1pt]
\end{tabular}
}
\label{tab:Robustness}
\end{table}

\subsubsection{Loss Function} 

\begin{table}[t]
\centering
\renewcommand{\arraystretch}{1.2}
\caption{Ablation results of transferability of HLIF for different fusion tasks.}
\Large
\resizebox{0.45\textwidth}{!}{
\begin{tabular}{lcccccc}
\toprule[1pt]

Config. & MSE$\downarrow$ & PSNR$\uparrow$ & SSIM$\uparrow$ & CC$\uparrow$ & Nabf$\downarrow$ & HyperIQA$\uparrow$ \\ 
 
\midrule \midrule
HoLoCo \cite{liu2023holoco} & 0.12 & 57.80 & 0.24 & 0.86 & 0.01 & 54.04 \\
\rowcolor[HTML]{CDE3FF}
+HLIF & \textbf{0.12} & \textbf{58.06} & \textbf{0.42} & 0.80 & \textbf{0.01} & \textbf{61.62} \\

LRRNet \cite{li2023lrrnet} & 0.07 & 59.93 & 0.33 & 0.62 & 0.04 & 34.54 \\
\rowcolor[HTML]{CDE3FF}
+HLIF & \textbf{0.03} & \textbf{64.22} & \textbf{0.51} & \textbf{0.66} & \textbf{0.01} & \textbf{39.29} \\

CoCoNet \cite{liu2024coconet} & 0.08 & 59.40 & 0.24 & 0.82 & 0.01 & 28.55 \\
\rowcolor[HTML]{CDE3FF}
+HLIF & \textbf{0.03} & \textbf{63.30} & \textbf{0.55} & \textbf{0.83} & \textbf{0.00} & \textbf{35.72}\\
\bottomrule[1pt]
\end{tabular}
}
\label{tab:Transferability}
\end{table}

To evaluate the properties of the proposed loss function, we study both the loss robustness of CDNet and the transferability of HLIF. Specifically, CDNet is trained with representative metric-based, perceptual, and gradient-based objectives, including SSIM-MSE \cite{raza2020pfaf, jian2020sedrfuse, liu2021learning, wang2022res2fusion, liu2024task}, the VGG perceptual loss used in LRRNet \cite{li2023lrrnet}, and the GIF loss from FFMEF \cite{zheng2023efficient}. As shown in Table~\ref{tab:Robustness}, CDNet remains effective under all tested losses, demonstrating its robustness to different supervision designs, while HLIF achieves the best overall performance. We further apply HLIF to representative baselines from different fusion tasks, including HoLoCo \cite{liu2023holoco}, LRRNet \cite{li2023lrrnet}, and CoCoNet \cite{liu2024coconet}. As shown in Table~\ref{tab:Transferability}, HLIF brings clear overall improvements, especially in terms of PSNR and Nabf, indicating that it not only provides the most suitable supervision for CDNet but also generalizes well as a plug-and-play loss.

\begin{table}[t]
\centering
\renewcommand{\arraystretch}{1.2}
\caption{Ablation results of training datasets on the SICE test set.}
\Large
\resizebox{0.45\textwidth}{!}{
\begin{tabular}{lcccccc}
\toprule[1pt]
Config. & MSE$\downarrow$ & PSNR$\uparrow$ & SSIM$\uparrow$ & CC$\uparrow$ & Nabf$\downarrow$ & HyperIQA$\uparrow$ \\
\midrule \midrule
MSRS \cite{tang2022piafusion} & 0.13 & 57.57 &  \cellcolor{bestblue}\textbf{0.45} & 0.88 & 0.01 & \cellcolor{secondblue}\underline{61.61} \\
LLVIP \cite{jia2021llvip} & \cellcolor{secondblue}\underline{0.12} & \cellcolor{secondblue}\underline{57.97} & 0.44 & \cellcolor{secondblue}\underline{0.88} & 0.01 &  \cellcolor{bestblue}\textbf{61.62} \\
SICE \cite{cai2018learning} &  \cellcolor{bestblue}\textbf{0.11} &  \cellcolor{bestblue}\textbf{58.28} & 0.43 &  \cellcolor{bestblue}\textbf{0.89} & \cellcolor{secondblue}\underline{0.01} & 60.89 \\
\bottomrule[1pt]
\end{tabular}
}
\label{tab:Datasets_supp}
\end{table}

\subsubsection{Training Datasets}

To evaluate whether the proposed framework depends heavily on specific training data distributions, we further compare models trained on MSRS, LLVIP, and SICE. As shown in Table~\ref{tab:Datasets_supp}, the model trained on the MSRS and LLVIP dataset still yields reasonable results on the SICE test set, suggesting that the joint design of CDNet and HLIF is not overly dependent on dataset-specific semantics.

\section{Conclusion}
\label{sec:conclusion}

In this paper, we propose CDNet, a novel and lightweight deep unfolding network for multi-source image fusion via combined dictionary learning. By integrating all source dictionaries into a unified update framework and performing fusion only on the luminance channel in the YCbCr color space, CDNet achieves effective fusion with only 6.5K parameters.
Complemented by the designed High- and Low-frequency Image Fidelity (HLIF) loss, which provides clear and modality-agnostic supervision, the method exhibits promising transferability across fusion scenarios. While only trained on the SICE dataset, CDNet generalizes robustly to diverse tasks including infrared-visible, medical image fusion, and fusion for semantic segmentation, without any task-specific fine-tuning. 
Comprehensive ablation studies further validate the effectiveness and robustness of CDNet, the transferability of HLIF, and the transferability of their joint design under different training datasets. Extensive experiments on benchmark datasets show that CDNet achieves strong fusion performance with high computational efficiency.
\Black{Future work will extend the combined dictionary unfolding framework for multi-modal perception and restoration, while preserving its joint-update efficiency.}

\bibliographystyle{IEEEtran}
\bibliography{mif}

\vfill

\end{document}